\documentclass[fleqn,10pt]{wlscirep}
\usepackage[T1]{fontenc} % For proper printing of special characters

\usepackage{threeparttable}
\usepackage{tabularx}
\usepackage{dcolumn}
\usepackage[title,page]{appendix}
\usepackage{subcaption}
\usepackage{stfloats}
\usepackage{float} % Required for [H]

\usepackage{xr}          % or xr-hyper if using hyperref
\usepackage[all]{hypcap}                            % For hyperref to point to the beginning of floats
\usepackage{comment}

\newcolumntype{d}[1]{D..{#1}}

\usepackage{adjustbox}

\makeatletter
\g@addto@macro{\UrlBreaks}{\UrlOrds}
\makeatother

\title{Navigating the Lobbying Landscape: Insights from Opinion Dynamics Models}

\author[1,*]{Daniele Giachini}
\author[1]{Leonardo Ciambezi}
\author[2]{Verdiana Del Rosso}
\author[2]{Fabrizio Fornari}
\author[3]{Valentina Pansanella}
\author[4,1]{Lilit Popoyan}
\author[5]{Alina Sîrbu}

\affil[1]{Sant'Anna School of Advanced Studies, Institute of Economics and Department L'EMbeDS, Pisa (PI), 56127, Italy} % Piazza Martiri della Libertà, 33 -
\affil[2]{University of Camerino, Department of Computer Science, Camerino (MC), 62032, Italy} % Via Madonna delle Carceri, 7 -
\affil[3]{National Research Council (CNR), Institute of Information Science and Technologies "A. Faedo" (ISTI), Pisa (PI), 56124, Italy} % G. Moruzzi 1 - 
\affil[4]{Queen Mary University of London, School of Business and Management, London, E14NS, United Kingdom} % 327 Mile End Rd, Bethnal Green - 
\affil[5]{University of Bologna, Department of Computer Science and Engineering, Bologna (BO), 40126, Italy} %  Mura Anteo Zamboni 7 -

\affil[*]{daniele.giachini@santannapisa.it}

\keywords{Opinion dynamics, Social networks, Agent-based modelling, Bayesian learning and behavioural bias, Lobbying, Climate change}

\begin{abstract}
While lobbying has been demonstrated to have an important effect on public opinion and policy making, existing models of opinion formation do not specifically include its effect. In this work we introduce a new model of lobbying-driven opinion influence within opinion dynamics, where lobbyists can implement complex strategies and are characterised by a finite budget. Individuals update their opinions through a learning process resembling Bayes-rule updating but using signals generated by the other agents (a form of social learning), modulated by under-reaction and confirmation bias. We study the model numerically and demonstrate rich dynamics both with and without lobbyists. In the presence of lobbying, we observe two regimes: one in which lobbyists can have full influence on the agent network, and another where the peer-effect generates polarisation. When lobbyists are symmetric, the lobbyist-influence regime is characterised by prolonged opinion oscillations. If lobbyists temporally differentiate their strategies, frontloading is advantageous in the peer-effect regime, whereas backloading is advantageous in the lobbyist-influence regime.
These rich dynamics pave the way for studying real lobbying strategies to validate the model in practice. 
\end{abstract}
\begin{document}

\flushbottom
\maketitle
\thispagestyle{empty}

\section*{Introduction}

Lobbying is a pervasive feature of modern policymaking, shaping decisions across virtually all domains of public life, from environmental regulation and healthcare to financial oversight and digital governance.  Beyond direct engagement with policymakers, lobbying strategies increasingly aim to influence public opinion, leveraging social networks and mass communication to create political momentum, reduce resistance to controversial policies, or shift the Overton window on key issues. Lobbyists, whether representing corporate interests, civil society, or ideological coalitions, act strategically to influence not only policymakers, but also public perception and discourse. A large body of empirical literature has demonstrated the effectiveness of lobbying in shaping policy outcomes \cite{grossman2001special,baumgartner2009lobbying}, with recent work underscoring the growing sophistication of lobbying strategies in networked political and media environments \cite{kluver2012informational,lewis2013advocacy}. This broader landscape of influence highlights the need to understand lobbying not only as a transactional activity, but also as a dynamic process of belief formation and diffusion in complex social systems.

Among the various mechanisms through which lobbying exerts influence, opinion formation within the public sphere plays a critical role. Lobbyists increasingly target public attitudes to create political pressure or legitimacy for specific policy choices \cite{schlichting2013strategic,stokes2017renewable}. The advent of digital platforms has magnified these efforts, enabling tailored messaging and decentralized opinion shaping. This makes the diffusion of opinions across social networks, where individuals interact, update beliefs, and potentially become advocates themselves, a crucial site for understanding lobbying efficacy.

To study this complex landscape, opinion dynamics models offer powerful tools. Opinion dynamics models are mathematical representations utilized to elucidate the intricate mechanisms underlying the dissemination and evolution of opinions or beliefs across heterogeneous populations \cite{sobkowicz2020whither,hegselmann2002opinion}. These models, rooted in theories of social influence and information diffusion, and also inspired by models from statistical physics, have garnered substantial attention and application across diverse disciplines, including sociology \cite{centola2007complex}, political science \cite{deffuant2002can}, marketing \cite{watts2007influentials} and economics \cite{bikhchandani1998learning}. They offer valuable insights into the emergence of collective behaviors, the formation of consensus, and the dynamics of information propagation within social networks \cite{das2014modeling,li2020effect, sirbu2017opinion}. Recently, they have also been used to model opinion formation in settings mediated by technology, for example through a bounded confidence model with algorithmic bias~\cite{sirbu2019algorithmic, pansanella2023mass}, which demonstrated how interaction through online social media that promotes information exchange with similar peers can cause fragmentation and polarization of opinions. 
In economic modelling and political science, opinion dynamics models have increasingly intersected with agent-based modeling approaches \cite{fagiolo2017macroeconomic,haldane2019drawing, bardoscia2025impact}, allowing for the incorporation of bounded rationality, motivated reasoning, and strategic behavior.

Despite their extensive utilization in numerous domains, the application of opinion dynamics models within the lobbying realm remains a nascent and underexplored area of inquiry. While existing research has shed light on the role of social networks and public discourse in shaping climate attitudes and policy preferences \cite{roser2018strategic}, the integration of formalized mathematical frameworks to analyze and forecast lobbying dynamics represents a promising frontier for advancing our understanding of the complex interplay between public opinion, lobbying discourse, and policy outcomes.
 However, relatively few models explicitly integrate lobbying agents into these frameworks. Many studies consider mass media influence in opinion dynamics, typically modelled as external field that can influence all agents with a certain probability \cite{das2014modeling,li2020effect, sirbu2017opinion,sirbu2013opinion, pansanella2023mass}. The effects observed depend on the structure of the population and model parameters: while in some situations mass media can stimulate consensus or speedup convergence, in other cases, especially when promoting extreme messages, it can facilitate polarisation. Alternatively, models can include zealots or stubborn agents \cite{mobilia2013commitment,sirbu2017opinion}, with similar effects. While some similarities exist, these approaches do not fully encompass the concept of lobbying, which can implement different dynamic strategies and typically have a cost. Furthermore, the modelling of opinions in these examples is very simplified, sometimes including bounded confidence or disagreement, but not taking into account more complex mechanisms such as confirmation bias or motivated reasoning. Therefore, the endogenous modeling of lobbying strategies aimed at belief manipulation within a dynamic opinion formation process remains underdeveloped.

 In this paper we focus on the opinion-formation layer of lobbying. In particular, how strategic actors shape public beliefs over time through advocacy, agenda setting and framing, and how grassroots/“outside” mobilization, alongside “inside” access strategies that complement opinion work. These channels are well documented in the political-economy and advocacy literatures and often operate precisely by shifting public attitudes to create political momentum or legitimacy for policy change. Our model abstracts from the policy decision stage and instead endogenizes lobbying-driven opinion influence as costly, time-staged communication under resource constraints. This scope is consistent with empirical evidence that frames and advocacy can materially affect public support and elite incentives \cite{grossman2001special,baumgartner2009lobbying,kluver2012informational,schlichting2013strategic,stokes2017renewable}.

This paper contributes to this growing interdisciplinary literature by introducing a novel model of lobbying influence over opinion dynamics in a boundedly rational social network. In our framework, a population of agents interacts repeatedly to update their subjective beliefs about the likelihood of an uncertain future event, for example, a damaging climate event. Each agent weighs two competing probabilistic models (one “optimistic” and one “pessimistic”) and updates their beliefs through a process resembling Bayes-rule updating but using signals generated by other agents (an extreme form of social learning \cite{DeGroot1974Consensus,Banerjee1992HerdBehavior,acemoglu2011opinion,Jadbabaie2012NonBayesian}), modulated by under-reaction \cite{epstein2010,massari2020,bottazzi2023} and confirmation bias (a form of directional motivated reasoning) \cite{druckman2019}. Agents are embedded in a directed network and interact through sequential signal exchanges that propagate beliefs over time. 
We then endogenize lobbying by introducing external agents (lobbyists) who strategically send costly signals to influence the population’s belief distribution by the end of a finite time horizon. Lobbyists differ in the model they support and are subject to a resource constraint (budget). Their objective is to minimize the distance between the final average belief and their preferred model. Lobbyists select randomized strategies from a constrained set of signalling paths, and their interventions shape both the speed and direction of belief evolution in the network.

Our numerical results reveal rich dynamics. In the absence of lobbying, the network tends to converge toward consensus or, under high confirmation bias, become polarized. A single lobbyist with sufficient resources can dominate belief formation and steer the network toward its preferred model, especially when agents are relatively open to new information. However, when two lobbyists with opposing goals act simultaneously, the system often fails to converge within the active time horizon, resulting in persistent instability and oscillation, especially in regions of the parameter space where both lobbyists are highly effective. We also examine the role of timing in lobbying strategies, showing that frontloading or backloading (i.e. concentrating signals at the beginning or at the end of the interaction time span, respectively) can be advantageous depending on the population’s biases. In the region of the parameter space where lobbying influence is stronger, backloading dominates, whereas in the region where peer effects dominate, frontloading prevails. Overall, our results indicate a complex interplay between cognitive constraints and strategic influence.

Our work relates to the literature on opinion dynamics with stubborn or exogenous influencers, including Acemoglu et al. (2013)\cite{acemouglu2013opinion}, that characterizes convergence and fluctuation patterns under linear averaging with stubborn nodes. Our framework differs along three dimensions. First, responsiveness is signal-dependent and state-dependent, capturing under-reaction and directional motivated reasoning (confirmation bias), rather than constant linear weights. Second, influential signals are endogenously generated by budget-constrained lobbyists, arriving as costly, time-staged messages that interact with peer signals. Third, the interaction of these features yields a regime structure with prolonged oscillations under symmetric lobbying which is distinct from fluctuation mechanisms in linear stubborn-agent models.

The remainder of the paper is structured as follows. The next section introduces the baseline opinion dynamics model, describing the agent interaction protocol and belief updating mechanism, and then incorporates lobbying agents and formalizes their strategic behavior, cost constraints, and impact on individual beliefs. The Results section presents the simulation results across three scenarios: no lobbying, single-lobbyist, and dual-lobbyist configurations. It explores the model's behavior across key parameters (under-reaction and confirmation bias), and performs sensitivity analysis on budget, network size, time horizon and lobbyist timing strategies. We then conclude by summarizing the findings and suggesting avenues for future research.

\section*{Methods: model description}
%\textcolor{red}{Topical subheadings are allowed. Authors must ensure that their Methods section includes adequate experimental and characterization data necessary for others in the field to reproduce their work.}

\subsection*{The baseline model}

We consider a social network of individuals who repeatedly interact communicating on the occurrence of an uncertain harmful event. For instance, they can do that by sharing posts, sending messages, and creating contents. We refer to any such item generically as a signal sent by an individual to their neighbors. For simplicity, signals are binary: they either assert that the event will occur or that it will not.
Guided by evidence of selective behavior in social networks\cite{Xiao2012}, individuals choose which type of signal to send based on their opinions. For example, an individual who assigns a high probability to the occurrence of the event will be more likely to share a signal affirming that it will occur. Consistent with evidence on social influence and contagion\cite{Muchnik2013SocialInfluenceBias,Bond2012FacebookMobilization,centola2007complex,AralWalker2011ViralDesignRCT}, received signals affect the recipients’ opinions.
Note that whether signals contain a kernel of truth is not essential for generating opinion change that, in turn, increases the likelihood of sharing similar signals. This is consistent with findings that false news diffuses more than true news on social media\cite{Vosoughi2018}. We model opinion formation and updating as a modification of Bayes’ rule: individuals behave as if signals from others are informative about the true data-generating process. At the same time, we allow the updating mechanism to deviate from exact Bayes to capture common processing biases, such as under-reaction\cite{Edwards_1982,epstein2010,massari2020,bottazzi2023} and confirmation bias\cite{nickerson1998confirmation,druckman2019}. This setup is also analogous to selective perception\cite{MassadHubbardNewtson1979}, a well-documented feature of social-media environments\cite{gearhart2020hostile,Xiao2012}.
Hence, our model represents an extreme form of social learning\cite{DeGroot1974Consensus,Banerjee1992HerdBehavior,acemoglu2011opinion,Jadbabaie2012NonBayesian}, in which individuals do not receive private signals from the true data-generating process but try to learn from what other agents share.

In formal terms, 
$N$ individuals interact for $\tau$ rounds about the occurrence of an uncertain damaging  
event after the interaction rounds. 
Individuals do not know the true probabilities that determine the occurrence of the event, but can rely on two probabilistic models.
Those are the probability distributions $(\pi_o,1-\pi_o)$ and $(\pi_p,1-\pi_p)$, where $\pi_k$, with $k\in\{o,p\}$, indicates the probability that the damaging event occurs. 
The model $(\pi_o,1-\pi_o)$ is \emph{optimistic} while the model $(\pi_p,1-\pi_p)$ is \emph{pessimistic}: $0<\pi_o<\pi_p<1$ with $\pi_o$ close to zero and $\pi_p$ close to one.

The social network is represented by a \emph{directed} graph, where a link from $i$ to $j$ means that agent $i$ communicates to agent $j$ but not viceversa (for example followers in online social networks).
We represent this by the $N\times N$ adjacency matrix $A=(a_{i,j})$. %Links are directed, thus, $a_{i,j}=1$ means that individual $i$ communicates to individual $j$, while $a_{i,j}=0$ means that $i$ does not communicate to $j$.  
Communication occurs in the interaction rounds. In each round $t\in\{1,\ldots,\tau\}$, an individual $\iota_t$ is independently and uniformly drawn from the population and it sends a signal $s_t\in\{0,1\}$ to the set of individuals to whom it communicates, that is $J(\iota_t)=\{j\in\{1,2,\ldots,N\}|a_{\iota_t,j}=1\}$. The signal $s_t=1$ indicates that the individual communicates that the event will occur, while $s_t=0$ indicates that the individual communicates that the event will not occur.

Each individual $i$ has a subjective probability distribution $\mathbf{p}_{i,t}=(p_{i,t},1-p_{i,t})$ on the realization of the event, with $p_{i,t}$ the probability attached at round $t$ to the case in which the event occurs. Such a distribution evolves according to the signals the individual receives from its network of connections and it is used to draw the signals it sends to the individuals to whom it communicates.
Concerning the evolution of subjective probabilities, for any $i\in\{1,2,\ldots,N\}$ and $t\in\{1,\ldots,\tau\}$, it is 
\begin{equation}    \label{eq:p}
    p_{i,t}=w_{i,t}\pi_o+(1-w_{i,t})\pi_p
\end{equation}
with $w_{i,t}\in(0,1)$. 
That is, any individual builds its subjective probabilities as the convex combination of the probabilistic predictions of the two models. The weights $w_{i,t}$ used to build subjective probabilities change depending on the received signals. In particular, for any individual $i$ and for any round $t$, the weight assigned to the optimistic model evolves according to
\begin{equation}
    w_{i,t}=\begin{cases}
    w_{i,t-1}    & \text{if } i\notin J(\iota_t),\\[0.5cm]
    \lambda_{i,t} w_{i,t-1}+(1-\lambda_{i,t})w_{i,t-1}\left(s_t\dfrac{\pi_o}{p_{i,t-1}}+(1-s_t)\dfrac{1-\pi_o}{1-p_{i,t-1}}\right)& \text{if } i\in J(\iota_t),
    \end{cases}
    \label{eq:w}
\end{equation}
where $\lambda_{i,t}\in[0,1]$ modulates the effects of cognitive biases of agent $i$ at round $t$ and $w_{i,0}\in(0,1)$ is the initial prior. 
As pointed out before, our agents build their beliefs in a way that resembles Bayesian updating but is applied in an extreme social-learning setting. They behave as if the observed social signals were i.i.d. draws from the true process, although these signals are noisy observations of other agents’ contemporaneous beliefs. In the absence of new evidence, and under common knowledge of identical $0.5$ priors, Bayesians would keep posteriors at $0.5$. By contrast, our agents treat social signals as informative about the state, departing from proper Bayesian-like learning. We nonetheless interpret $\lambda_{i,t}$ as capturing specific cognitive biases on top of this process. In particular, although proper Bayesian learning cannot be recovered in this social-learning environment, setting $\lambda_{i,t}=0$ makes agent $i$ update by Bayes’ rule and this acts as a benchmark within the proposed setting.
If $\lambda_{i,t}=1$ then $i$ keeps the initial combination fixed no matter the sequence of signals it receives. Intermediate values of $\lambda_{i,t}$ indicate that, as $i$ receives a signal, it updates the weights in the right direction but in lower magnitude than what Bayes' rule would prescribe. Thus, individuals can underreact to signals \cite{epstein2010,massari2020,bottazzi2023}, producing the well-documented pattern in belief updating often referred to as conservatism\cite{Edwards_1982}.
We enrich such a behavioural interpretation imposing that $\lambda_{i,t}$ depends upon the type of signal received and the prior. In this way, it captures a form of directional motivated reasoning that, in turn, gives rise to confirmation bias (the tendency of individuals to discard information that contradicts their priors), another common feature in the formation of %climate-related 
beliefs \cite{nickerson1998confirmation,druckman2019}. In particular, we assume
\begin{equation}
\label{eq:lambda}
    \lambda_{i,t}=\phi_i|1-s_t-w_{i,t-1}|+(1-\phi_i)\lambda_{i}
\end{equation}
with $\phi_i,\lambda_i\in[0,1]$. Note that, our updating rule differs from linear averaging rules commonly studied in stubborn-agent models\cite{acemouglu2013opinion}, because the effective step size is endogenous to the signal’s (in)congruence with priors (equation \eqref{eq:lambda}). This feature is central to the regime behavior we document later. 

The overall degree of under-reaction is the convex combination of two components. On the one hand, the parameter $\lambda_i$ captures the baseline level of under-reaction of agent $i$. On the other hand, the function $|1-s_t-w_{i,t-1}|$ introduces directional motivated reasoning or, equivalently for our analysis and interpretation, confirmation bias. The parameter $\phi_i$ regulates the strength of the directional motivated reasoning. Suppose, for instance, that directional motivated reasoning is strong ($\phi_i\simeq1$), if an individual has a strong prior in the optimistic model ($w_{i,t-1}\simeq1$) and receives a signal that favours it with respect to the pessimistic model ($s_t=0$) then its overall degree of under-reaction is low: information has a sensible impact on its beliefs. If, instead, the information contradicts its priors ($s_t=1$), then the degree of under-reaction is high: information has a negligible effect on its beliefs. A symmetric argument holds if the agent's priors favour the pessimistic model. If, instead, directional motivated reasoning is weak ($\phi_i\simeq0$), then the individual under-reacts with respect to information because of the effect of $\lambda_i$, but the nature of the signal and its relation to the prior have a negligible effect on probability updating. As mentioned above, this setup is also akin to a form of selective perception.

Selective behavior in the generation of signals is modelled assuming that the signal at round $t$ is drawn from the distribution ${\bf p}_{\iota_t,t}$ (i.e., $\text{Prob}\{s_t=1\}=p_{\iota_t,t}$). Thus, the timeline of events in each round $t$ is as follows: $i)$ $\iota_t$ is independently and uniformly drawn from the population; $ii)$ the set of receivers $J(\iota_t)$ is created; $iii)$ the signal $s_t$ is drawn from the distribution ${\bf p}_{\iota_t,t}$; $iv)$ individual weights are updated according to equation \eqref{eq:w}.

%Assuming that the adjacency matrix of the network of agents is randomly drawn, the belief distribution in the baseline model turns out to be characterised by path-dependence. Indeed, depending on the particular sequence of random draws, one may observe different scenarios. 
%In Figure \ref{fig:ex_base} one can observe how, under the same set of parameter values, changing the seed of the random number generator causes two qualitatively opposite scenarios. In one case polarization around the optimist model emerges, while, in the other, polarization around the pessimist model is observed.

%\begin{figure}[t]
%\center
%  \includegraphics[width=0.49\linewidth]{figures_clipped/baseline.png}
%  \includegraphics[width=0.49\linewidth]{figures_clipped/baseline1.png}
%  \caption{Two examples of belief distribution with $N=1,\!000$, $T=100$, $\pi_o=0.01$, $\pi_p=0.99$, $\lambda_{i,t}=0.8\;\forall i$, random adjacency matrix, and different seeds. \textcolor{red}{TO BE UPDATED: il sistema dei lambda è cambiato rispetto a queste immagini. Andrebbero quantomeno aggiornate.}}
%  \label{fig:ex_base}
%\end{figure}

\subsection*{Introducing lobbyists} \label{Sec:LobbyModel}
Lobbyists can be considered external agents with respect to the social network that, in each turn, can pay a cost to send signals to individuals. Their objective is to influence the time $\tau$ distribution of beliefs so that it becomes concentrated on one of the two models. We emphasize that this module captures the opinion-formation channel of lobbying; the policy-adoption stage is not modelled and is treated as exogenous to the opinion dynamics studied here.

Assume that there are $L$ active lobbyists and define $S_{i,t}^\ell\in\{0,1\}$ as a variable indicating whether lobbyist $\ell\in\{1,2,\ldots,L\}$ sends a signal to individual $i$ at the beginning of round $t$ ($S^\ell_{i,t}=1$) or not ($S^\ell_{i,t}=0$). 
It follows that a pure strategy for lobbyist $\ell$ can be indicated as a matrix ${S}^\ell\in\mathcal{S}=\{0,1\}^{N\times \tau}$. 
Each lobbyist supports one model between the optimist and the pessimist one and the signals it sends are in favour of the model it supports. For instance, a lobbyist supporting the optimistic model will signal 0, that is, the event is not occurring. 
Define the function that assigns to each lobbyist the model it supports as $m: \{1,2,\ldots,L\}\to\{o,p\},\ell\mapsto m(\ell)$.  We assume that signals are costly. In particular, the cost of a signal is fixed to a given level $c>0$ and each lobbyist $\ell$ has a budget $B^\ell$. 
Without loss of generality, we shall set $c=1$ in what follows. That is,  $B^\ell$ shall be understood as the maximum number of signals lobbyist $\ell$ can send. Notice that this assumption constrains the set of strategies that each lobbyist can actually play, that is, a feasible strategy is such that
$\sum_{t=1}^\tau\sum_{i=1}^N \, S^\ell_{i,t}\leq B^\ell$.
Hence, we define the set of feasible strategies for lobbyist $\ell$ as $\mathcal{S}_\ell=\{S^\ell\in\mathcal{S}\,|\,\sum_{t=1}^\tau\sum_{i=1}^N \, S^\ell_{i,t}\leq B^\ell\}$.
We allow lobbyists to randomize their choices. Call $\Delta(\mathcal{S}_\ell)$ the mixed extension of $\mathcal{S}_\ell$ and  $\sigma^\ell({S})$ the probability that lobbyist $\ell$ attaches to a feasible strategy ${S}$, such that $\sigma^\ell\in\Delta(\mathcal{S}_\ell)$ is a feasible mixed strategy for lobbyist $\ell$. Note that, this strategic, budget-constrained sending of costly, time-staged signals differs from exogenous stubborn-node formulations. Indeed, each individual reacts to the signal received in the same way as they do for peer signals. Therefore, now, individual probabilities also evolve as a consequence of the strategies of the lobbyists. 
In particular, call $S^{-\ell}=(S^1,\ldots,S^{\ell-1},S^{\ell+1},\ldots,S^L)\in\mathcal{S}_1\times\ldots\times \mathcal{S}_{\ell-1}\times\mathcal{S}_{\ell+1}\times\ldots\times\mathcal{S}_L$ a profile of feasible pure strategies for the lobbyists different from $\ell$, such that $p_{i,\tau}(S^\ell,S^{-\ell})$ indicates the final probability of individual $i$ as a function of the strategies played by all lobbyists. 
Define the vector of lobbyists' signals received by agent $i$ at time $t$ as $\mathcal{L}^{i,t}\in\{0,1,2\}^L$, where $\mathcal{L}^{i,t}_\ell=0$ if $S^\ell_{i,t}=0$, $\mathcal{L}^{i,t}_\ell=1$ if $S^\ell_{i,t}=1$ and $m(\ell)=p$, $\mathcal{L}^{i,t}_\ell=2$ if $S^\ell_{i,t}=1$ and $m(\ell)=o$. Thus, assume that lobbyists send signals at the beginning of each round $t$ and that the order in which their signals are received by individuals is $Z_t\in\text{Perm}\{1,\ldots,L\}$, i.e. an element of the set of all permutations of the indexes of lobbyists (in what follows, we assume that the order according to which lobbyists send their signals is randomly chosen). The weight assigned to the optimistic model deriving from previous interaction needs to be updated for any $z\in \{1,\ldots,L\}$ according to
\begin{equation}
    w^{z}_{i,t-1}=\begin{cases}
    w^{z-1}_{i,t-1}    & \text{if } \mathcal{L}^{i,t}_{\ell_z}=0,\\[0.5cm]
    \lambda^{1,z}_{i,t} w^{z-1}_{i,t-1}+(1-\lambda^{1,z}_{i,t})w^{z-1}_{i,t-1}\dfrac{\pi_o}{\pi_o w^{z-1}_{i,t-1}+\pi_p(1-w^{z-1}_{i,t-1})}& \text{if } \mathcal{L}^{i,t}_{\ell_z}=1,\\[0.5cm]
    \lambda^{2,z}_{i,t} w^{z-1}_{i,t-1}+(1-\lambda^{2,z}_{i,t})w^{z-1}_{i,t-1}\dfrac{1-\pi_o}{1-\pi_o w^{z-1}_{i,t-1}-\pi_p(1-w^{z-1}_{i,t-1})}& \text{if } \mathcal{L}^{i,t}_{\ell_z}=2,
    \end{cases}
    \label{eq:wl}
\end{equation}
where $\ell_z$ indicates the lobbyist index number appearing in $z$-th position of the $Z_t$ vector,
$w^0_{i,t-1}=w_{i,t-1}$, $\lambda^{1,z}_{i,t}=\phi_iw^{{z-1}}_{i,t-1}+(1-\phi_i)\lambda_i$, and $\lambda^{2,{z}}_{i,t}=\phi_i(1-w^{z-1}_{i,t-1})+(1-\phi_i)\lambda_i$. In this way, the cognitive biases discussed in advance influence belief updating also when signals are sent by lobbyists. The weight updating rule for the signals received from other individuals needs to be adjusted, indeed, equation \eqref{eq:w} now becomes
\begin{equation}
    w_{i,t}=\begin{cases}
    w^L_{i,t-1}    & \text{if } i\notin J(\iota_t),\\[0.5cm]
    \begin{aligned}
    \lambda_{i,t} w^L_{i,t-1}+(1-\lambda_{i,t})w^L_{i,t-1}\left(s_t\dfrac{\pi_o}{\pi_o w^{L}_{i,t-1}+\pi_p(1-w^{L}_{i,t-1})}\right.
    \left.+(1-s_t)\dfrac{1-\pi_o}{1-\pi_o w^{L}_{i,t-1}-\pi_p(1-w^{L}_{i,t-1})}\right) 
    \end{aligned}
    & \text{if } i\in J(\iota_t),
    \end{cases}
    \label{eq:w1}
\end{equation}
with $\lambda_{i,t}=\phi_i|1-s_t-w^L_{i,t-1}|+(1-\phi_i)\lambda_i$ as in advance.
Subjective probabilities, instead, remain as in equation \eqref{eq:p} since they are computed at the end of the interaction round.  

Finally, we assume that the payoff of lobbyist $\ell$ using the feasible mixed strategy $\sigma^\ell$ when the other lobbyists use the feasible mixed strategies $\sigma^1,\ldots,\sigma^{\ell-1},\sigma^{\ell+1},\ldots,\sigma^L$ is 
\begin{equation*}
U_\ell=-\text{E}\left[\,
\sum_{S^1\in\mathcal{S}_1}\cdots\sum_{S^L\in\mathcal{S}_L}\;\prod_{l=1}^L\sigma^l(S^l) \left|\frac{1}{N}\sum\limits_{i=1}^N  p_{i,T}(S^\ell,S^{-\ell})-\pi_{m(\ell)}\right|\,\right]\,,
\label{eq:lob_prob}
\end{equation*}
where the expectation $\text{E}$ is computed with respect to the random variables deciding the order in which lobbyists send their signals, the selection of communicating individuals, and the signalling choices operated by individuals.
The underlying intuition is that a lobbyist benefits as the expected distance between the average final belief and the model it supports decreases.

When lobbyists are active, the timeline of events in each round $t$ becomes: $i)$ each lobbyist sends the signals to the individuals it selected; $ii)$ individual weights are updated according to equation \eqref{eq:wl}; $iii)$ $\iota_t$ is independently and uniformly drawn from the population; $iv)$ the set of receivers $J(\iota_t)$ is created; $v)$ the signal $s_t$ is drawn from the distribution ${\bf p}_{\iota_t,t}$; $vi)$ individual weights are updated according to equation \eqref{eq:w1}.

\subsubsection*{Example 1: one individual and one rational lobbyist}
A simple illustrative example is the special case in which $N=L=1$. Assume that the lobbyist supports the pessimistic model and that $w_{i,0},\lambda_1,\phi_1\in(0,1)$. By direct inspection of equation \eqref{eq:wl}, one notices that the weight assigned to the optimistic model decreases  if $S^1_{1,t}=1$ (for any $t$). This implies that the payoff of the lobbyist increases for each signal sent. Thus, a lobbyist that wants to maximize its payoff will use all of its budget to send signals. Since the specific timing of the signals does not matter, the optimal strategy of the lobbyist is 
\begin{equation*}
    S^{1,*}=
    \begin{cases}
        (1,\ldots,1) & \text{if } \tau\leq B^1,\\
        S^1\in\{(S^1_{1,1},\ldots,S^1_{1,\tau})|\sum_{t=1}^\tau S^1_{1,t}=\lfloor B^1\rfloor \} & \text{if } \tau> B^1.
    \end{cases}
\end{equation*} 
\subsubsection*{Example 2: two unbiased individuals, one period, and two rational lobbyists with unitary budget supporting opposite models}
The second special case we consider is characterized by $N=L=2$, $\tau=1$, $B^1=B^2=1$, and $\lambda_i=\phi_i=0\;\forall i\in\{1,2\}$. Without loss of generality, we assume that lobbyist 1 supports the pessimistic model while lobbyist 2 supports the optimistic model.
Since both time and budget are unitary (and not sending signals is strictly dominated), each lobbyist has two feasible pure strategies: sending a signal to individual 1 or to individual 2, that is, $S^\ell\in\mathcal{S}_\ell=\{(0,1),(1,0)\}$. Under the assumptions that the model structure is common knowledge and the lobbyists are rational, we can compute the Nash equilibrium \cite{nash1950} resulting from the strategic interaction of the two lobbyists.

Notice that with $\lambda_i=\phi_i=0\;\forall i\in\{1,2\}$ individuals follow Bayes-rule updating under the extreme social learning framework that characterizes our model. In this one-period, signal-symmetric case, the order of receipt is immaterial for the final posteriors. Thus, to compute final probabilities, we only need to consider the following set of possible received signals: 
\begin{equation*}
\Omega=\left\{\emptyset,\{0\},\{1\},\{0,0\},\{0,1\},\{1,1\},\{0,0,1\},\{0,1,1\}\right\}.
\end{equation*}
That is, since the lobbyists support opposite models and one of the individuals sends one signal to the other, the most extreme cases are the one in which an individual does not receive any signal, and the one in which it receives two (different) signals from the lobbyists and a signal from the other agent. All the other cases in between are possible. Assuming a uniform prior distribution for both individuals, the two agents are \emph{ex-ante} identical, meaning that the differences in their final probabilities are uniquely generated by the signals they received. Hence, we can simplify the notation calling $w(\omega)$ the weight assigned to the optimistic model by an agent that has received the signals $\omega\in\Omega$. As a direct consequence, we define the probability such an agent assigns to the realization of the event as $p(\omega)= w(\omega)\pi_o+(1-w(\omega))\pi_p$.   
Then, calling $t_s$, with $s\in\{0,1\}$, the number of times $s$ occurs in $\omega$, one has
\begin{equation*}
    w(\omega)=\frac{\pi_o^{t_1}(1-\pi_o)^{t_0}}{\pi_o^{t_1}(1-\pi_o)^{t_0}+\pi_p^{t_1}(1-\pi_p)^{t_0}}\,.
\end{equation*}
Given the previous assumptions, the symmetry they entail, and the fact that $p(\omega)\in(\pi_o,\pi_p)\;\forall \omega$, the payoff of lobbyist  $\ell\in\{1,2\}$ conditional upon both lobbyists playing the same pure strategy ($S^1=S^2$, lobbyists send their signals to the same individual) is 
\begin{equation*}
    u_\ell(S^1=S^2)=-\left|\frac{1}{4}\left(p(0,1)(1+p(1))+(1-p(0,1))p(0)+p(\emptyset)(1+p(0,1,1))+(1-p(\emptyset))p(0,0,1)\right)-\pi_{m(\ell)}\right|\,,
\end{equation*}
while the payoff of $\ell$ conditional upon the two lobbyists choosing different pure strategies ($S^1\ne S^2$, lobbyists send their signals to different individuals) is
\begin{equation*}
    u_\ell(S^1\ne S^2)=-\left|\frac{1}{4}\left(p(1)(1+p(0,1))+(1-p(1))p(0,0)+p(0)(1+p(1,1))+(1-p(0))p(0,1)\right)-\pi_{m(\ell)}\right|\,.
\end{equation*}
Thus, for a generic mixed strategy profile, the payoff of lobbyist $\ell$ reads 
$$U_\ell=(\sigma^1(1,0)\sigma^2(1,0)+\sigma^1(0,1)\sigma^2(0,1))u_\ell(S^1=S^2)+(\sigma^1(0,1)\sigma^2(1,0)+\sigma^1(1,0)\sigma^2(0,1))u_\ell(S^1\ne S^2)\,.$$
However, notice that $u_1(S^1=S^2)-u_1(S^1\ne S^2)=u_2(S^1\ne S^2)-u_2(S^1= S^2)$, hence, neglecting the trivial case in which the differences are zero, there generically exists a unique Nash equilibrium characterised by $\sigma^\ell(1,0)=\sigma^\ell(0,1)=1/2\;\forall \ell \in\{1,2\}$.

\subsection*{Simulation setup}

In all of our simulations, we consider a complete network of 500 agents. The adjacency matrix $A$ is therefore of size $500 \times 500 $, with diagonal entries $a_{i,i} = 0$, and off-diagonal entries $a_{i,j \neq i} = 1$. Each agent is characterised by an individual-specific initial weight $w_{i,0}$ which is randomly drawn from an uniform distribution with support $(0,1)$.
The probabilities characterizing the two models (optimistic and pessimistic) %to the "climate change event" 
are $\pi_{o} = 0.01$ and $\pi_{p} = 0.99$.
We consider three main scenarios.
In the ``baseline'' scenario, the network is left to interact with itself, with no outside interference, i.e. no lobbyist. In the ``one-lobbyist'' scenario, we introduce a single active lobbyist ($L = 1$) which supports the pessimistic model ($m(1) = p$). In the ``two-lobbyists scenario'', we introduce two active lobbyists that support competing models ($m( 1) = p$, and $m( 2) = o$).
In each scenario contemplating lobbyists, each one of them is endowed with a budget $B = 10,\!000$.
For the purpose of our numerical analysis, we introduce the parameter $T\leq\tau$, representing the lobbyists' time horizon, that is, the time span over which it allocates its budget $B^\ell$. Accordingly, we set $S^\ell_{i,t}=0\;\forall t\in(T,\tau]$. This constraint allows us to manage the potentially vast space of feasible lobbying strategies and enables analysis in scenarios with large $\tau$. In particular, neglecting the zeros appearing after $T$,  we assume that lobbyists dispose of a set of 100 strategy matrices of size $(T \times 500)$ in which the variable indicators for signals are distributed randomly over nodes and iterations, and choose one of those matrices at random at the beginning of each independent simulation run. For most of our experiments, we set $T=100$. This means that, over 100 periods, each lobbyist is able to send 100 signals per period, thus reaching out to 20\% of the nodes at each time-step. Moreover, this strategy selection procedure resembles an adaptation of the uniform-probability mixed strategy that emerges as the Nash equilibrium in Example 2.  Throughout, we interpret results as effects on lobbying-driven opinion influence; any policy-adoption mapping is intentionally left outside the model.

\begin{table}[ht!]
\centering
\begin{tabular}{llc}
\hline
\textbf{Parameter} & \textbf{Description} & \textbf{Baseline value}  \\
\hline
$N$ & Number of individuals in the network & 500 \\
$T$ & Number of interaction rounds & 100  \\
$\pi_o$ & Probability of damaging event under optimistic model & 0.01 \\
$\pi_p$ & Probability of damaging event under pessimistic model & 0.99 \\
$\phi$ & Strength of directional motivated reasoning/confirmation bias & $[0,1]$  \\
$\lambda$ & Degree of baseline under-reaction & $[0,1]$  \\
$B$ & Budget of each lobbyist & 10000  \\
$c$ & Cost per signal sent by each lobbyist & 1  \\
\hline
\end{tabular}
\caption{Summary of model's parameters, their descriptions, and their baseline numerical values.}

\label{tab:model_parameters}
\end{table}

As shown in equation \eqref{eq:lambda}, each $\lambda_{i,t}$ is subject to two different types of bias, captured by the parameters {$\lambda_i$ and $\phi_i$}. While $\lambda_i$ captures the degree of under-reaction to \emph{any} new information, $\phi_i$ represents the strength of directional motivated reasoning (or confirmation bias), which is signal-dependent. 
We explore the properties of the model assuming homogeneous biases ($\lambda_i=\lambda$ and $\phi_i=\phi$ for all individuals $i$) and considering different configurations of the network with respect to these two independent dimensions, i.e. a parameter space generated by $(\lambda, \phi) \in [0,1]\times [0,1]$. For reference, Table \ref{tab:model_parameters} summarizes the model parameters, their descriptions, and the baseline values.

In all simulations, we monitor various criteria. An important aspect is the number of clusters that form in the population: consensus means one cluster, more clusters mean fragmentation and polarisation. We employ the effective number of clusters defined as $C={N^2}/{\sum_i^k N_i^2 }$, where $k$ is the number of clusters and $N_i$ is the size of each cluster. This measure takes into account not only the number of groups, but also their size. $C$ is equal to $k$ when we have $k$ clusters of equal size, and it is smaller when the clusters are imbalanced. A second criterion of interest is the average opinion of agents at the end of the simulation, which we report in terms of average of the $p_{i,\tau}$ values as defined in equation \eqref{eq:p}. In addition to indicating how beliefs have collectively shifted, this criterion is also informative of lobbyists' payoff.
Furthermore, we look at the number of iterations required for convergence, indicating the speed of opinion cluster formation. For all parameter configurations we provide mean values of these criteria over 150 independent runs,  each one of which is simulated until the model reaches convergence (i.e., until all the individual weights assigned to the competing models stop updating). 

\section*{Results}
\label{Sec:Results}

\subsection*{Baseline scenario}

\begin{figure}[ht!]
\center
  \begin{subfigure}[b]{0.45\textwidth}
        \includegraphics[width=\linewidth]{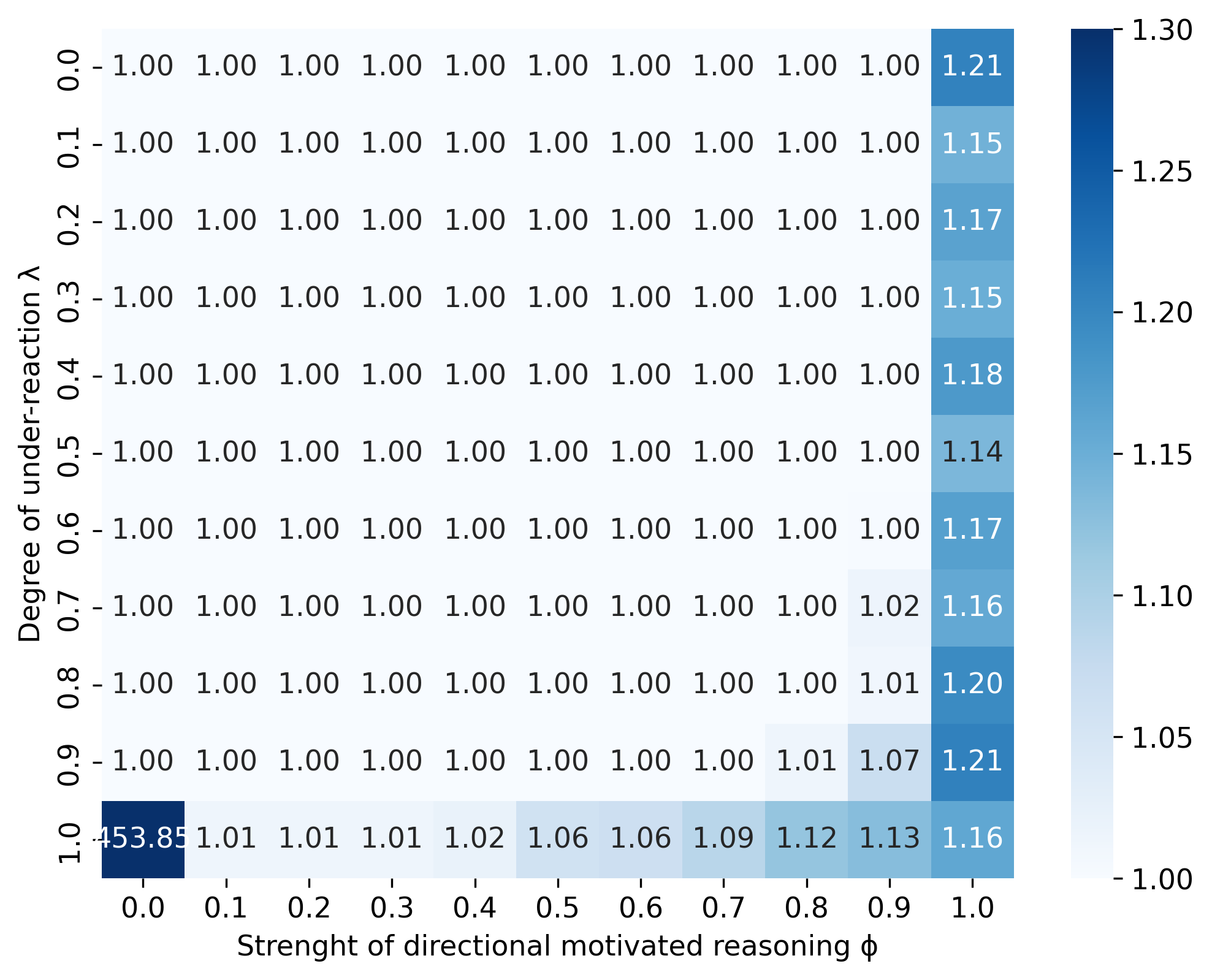}
        \caption{Number of clusters: whole parameter space}
    \end{subfigure}
    \hspace{0.05\textwidth}
    \begin{subfigure}[b]{0.45\textwidth}
        \includegraphics[width=\linewidth]{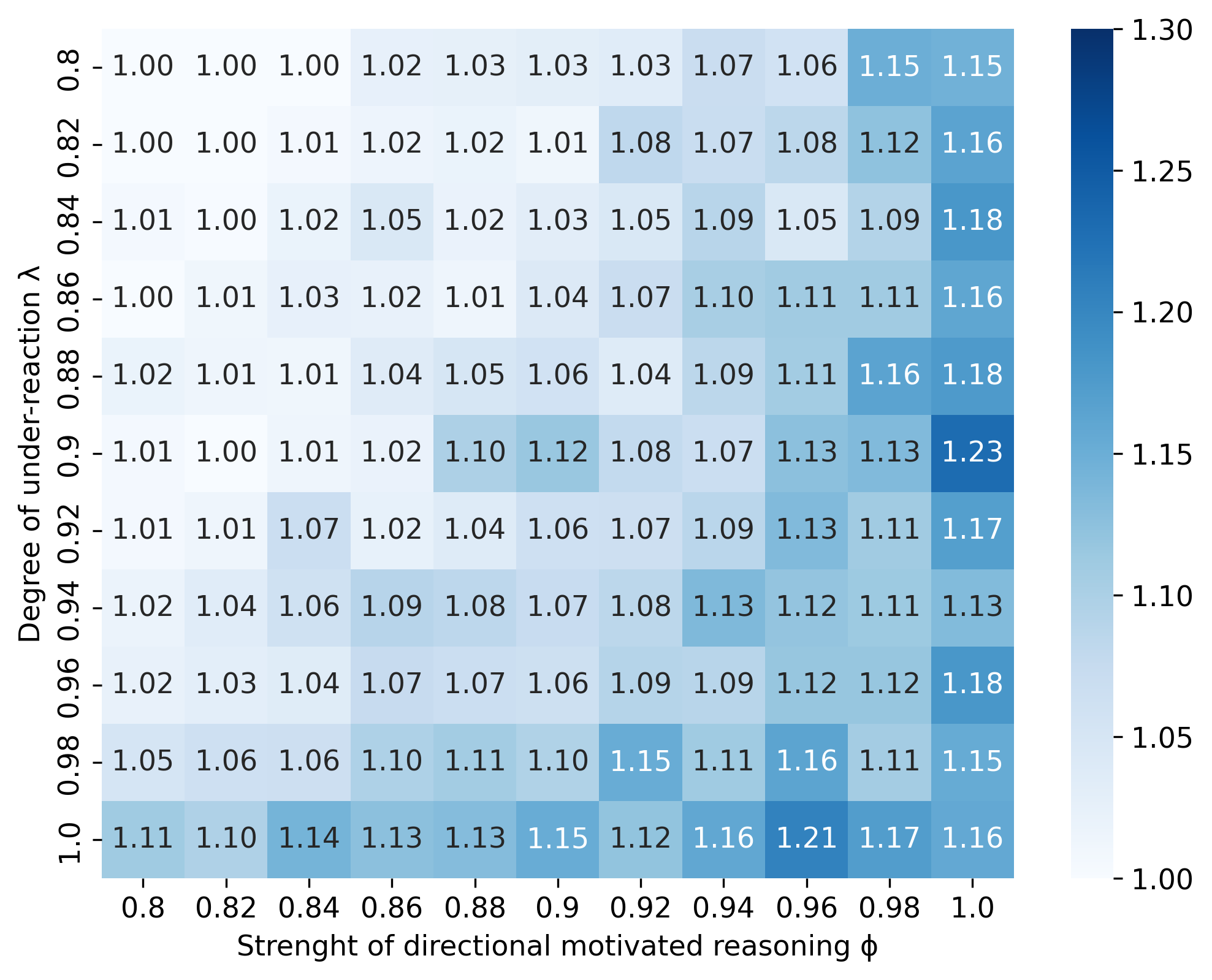}
        \caption{Number of clusters: bottom-right corner zoom}
    \end{subfigure}
    \begin{subfigure}[b]{0.45\textwidth}
        \includegraphics[width=\linewidth]{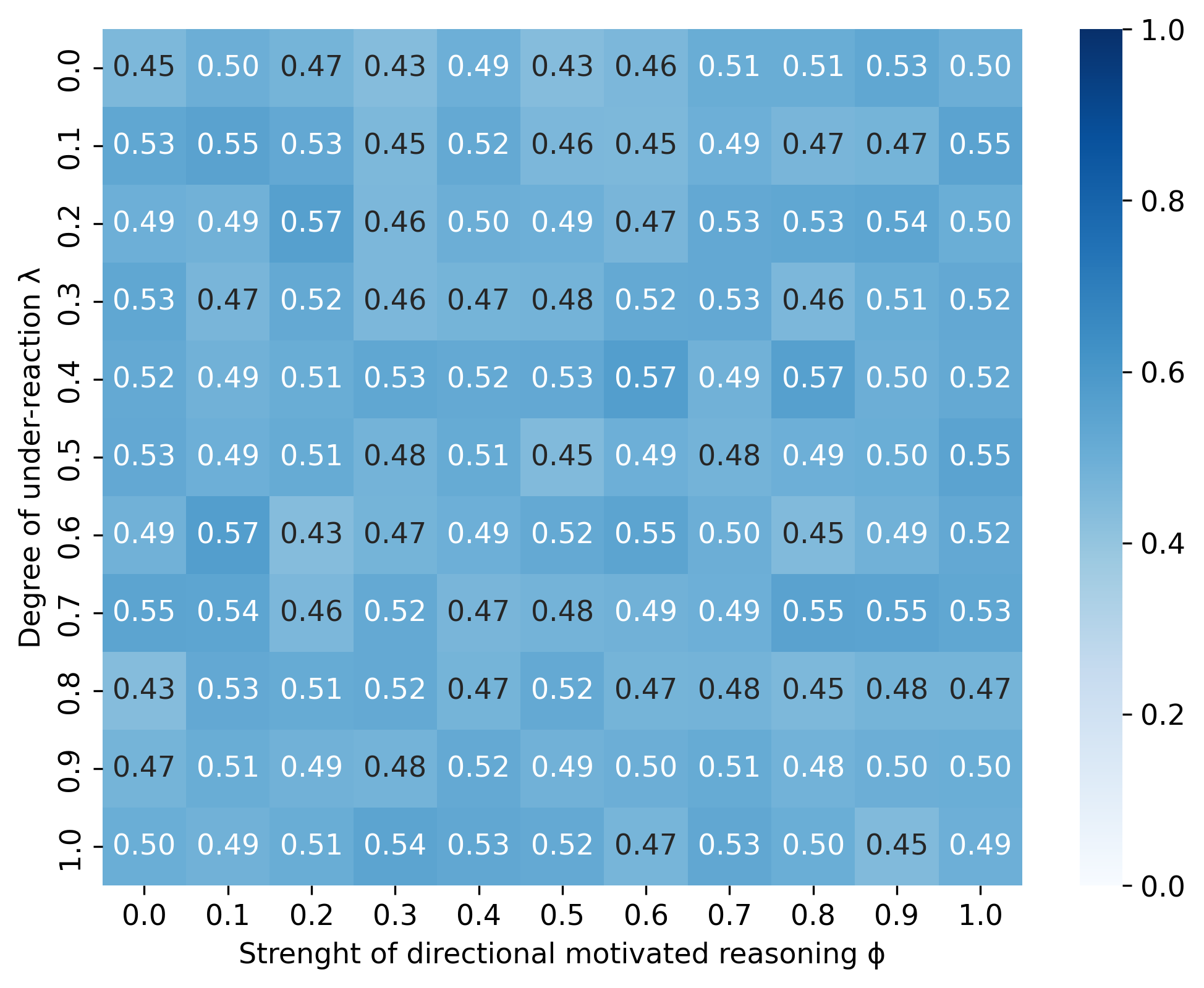}
        \caption{Average subjective probabilities: whole parameter space}
    \end{subfigure}
    \hspace{0.05\textwidth}
     \begin{subfigure}[b]{0.45\textwidth}
        \includegraphics[width=\linewidth]{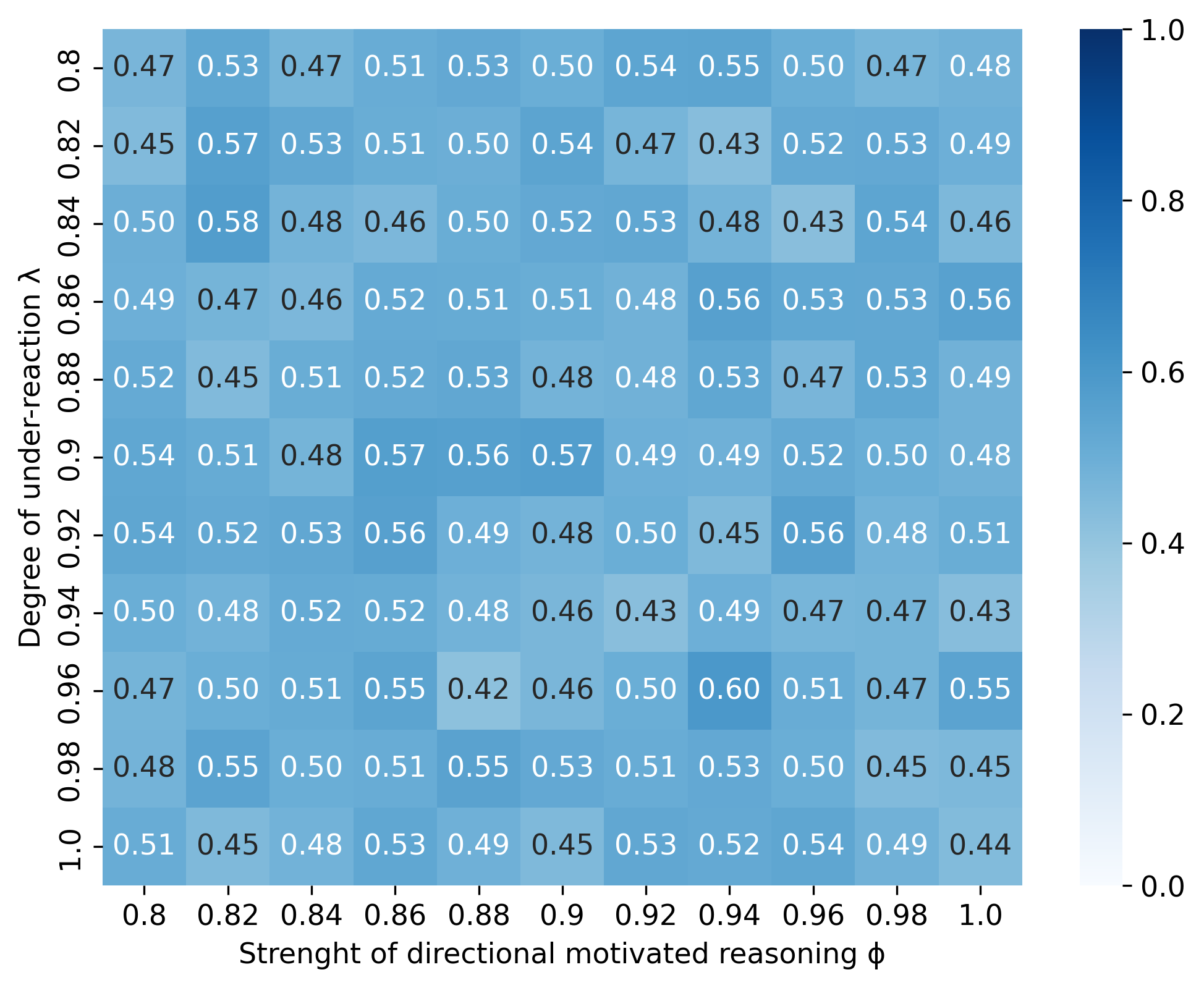}
        \caption{Average subjective probabilities: bottom-right corner zoom}
    \end{subfigure}
  \caption{Baseline (no-lobbyist) scenario. The average effective number of clusters and the final average subjective probabilities of agents are represented as a function of strength of directional motivated reason $\phi$ and the degree of under-reaction $\lambda$. %Values are averaged on 150 independent runs of each setting. 
  In the case where $\lambda=1.0$, $\phi = 0.0$, the number of clusters metric is not applicable, since the agents do not change their opinions over time from their initial conditions, randomly drawn from an uniform distribution in [0,1].}  
  \label{fig:heatmap_cluster0}
\end{figure}

\begin{figure}[ht]
\center
\begin{subfigure}[b]{0.45\textwidth}
 \includegraphics[width=\linewidth]{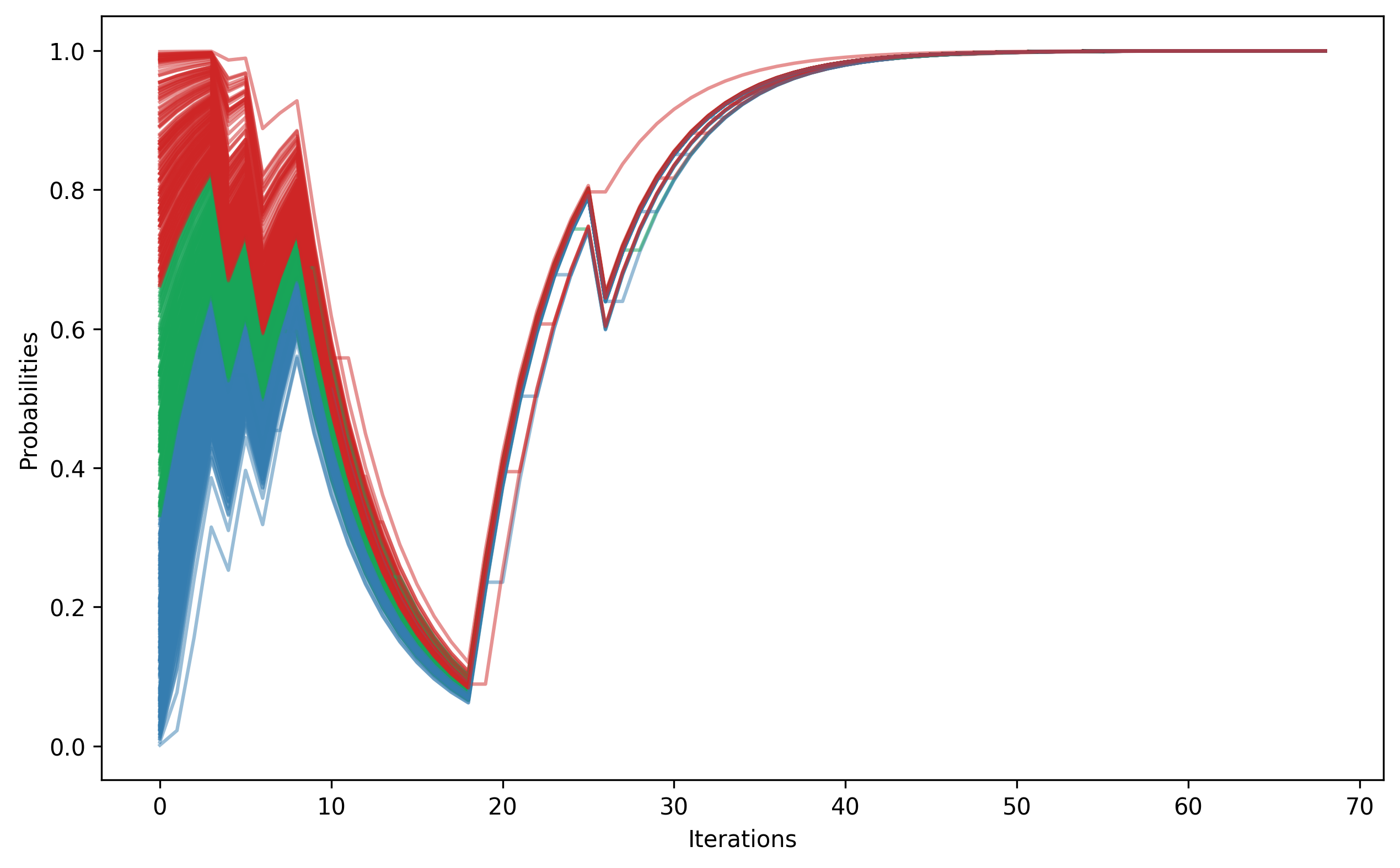}
 \caption{Convergence to one cluster. Each agent has a degree of under-reaction   $\lambda=0.8$ and a strength of directional reasoning $\phi = 0.0$. After about 60 time steps, all agents of the network have a final subjective probability $p_i = 0.99$. }
 \label{fig:evolution_1}
\end{subfigure}
\begin{subfigure}[b]{0.45\textwidth}
 \includegraphics[width=\linewidth]{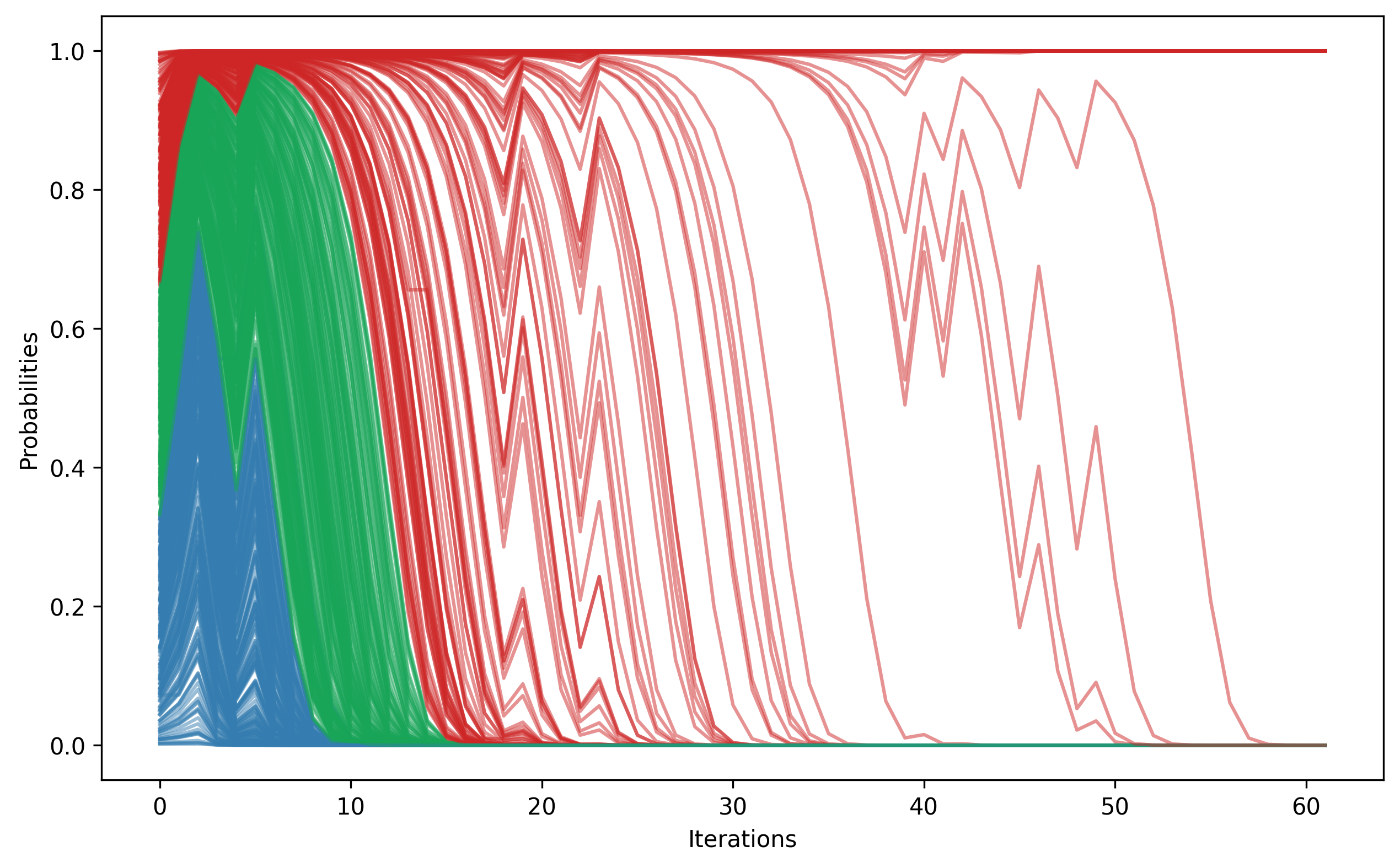}
 \caption{Convergence to two clusters. Each agent has a degree of under-reaction  $\lambda=1.0$ and a strength of directional reasoning $\phi = 0.9$. After about 60 iterations, a cluster of agents of the network ($\approx16\%$) has a final subjective probability $p = 0.99$, while the rest of the agents (84\%) has the probability $p = 0.01$.}
 \label{fig:evolution_2}
\end{subfigure}
  \caption{Opinion evolution. Examples of the evolution of the agent's subjective probabilities in the baseline scenario with and without final consensus of the network. %The simulations are realized with a fully connected network with $N=500$ agents, optimistic and pessimistic models probabilities equals to $\pi_o=0.01$ and $\pi_p=0.99$, respectively. 
  }
\end{figure}

In the first setting, we investigated the properties of the model in the absence of any active lobbyist, when the nodes are left free to interact with their peers without outside input.
Figure \ref{fig:heatmap_cluster0} displays the effective number of clusters and the average subjective probabilities across the parameter space when the stable state is reached. We observe that, for most of the parameter region under consideration, the distribution of subjective probabilities tends to converge to a general consensus, with all of the nodes eventually supporting a single model (either $o$ or $p$, depending on initial conditions). This dynamic is relatively quick and is also illustrated by the evolution plot in Figure \ref{fig:evolution_1}. 

Two notable exceptions to this behavior are worth mentioning: the first, and more trivial, occurs in the special case where $\lambda = 1$ and $\phi = 0$. In this case, as it is immediately clear from equation \eqref{eq:lambda}, nodes are ``deaf'' to whatever signal they receive from their peers, and any update of subjective probabilities is impossible. In this case, there is no dynamics and each agent maintains its initial condition.
The second exception emerges at the bottom and to the right of the parametric region, where the confirmation bias parameter $\phi$ is sufficiently large to prevent the network from reaching a consensus.
In this case, in many runs the model converges to a polarized equilibrium, in which a variable proportion of agents supports one model, while the rest of the network supports the competing one (see Figure \ref{fig:evolution_2} for an example). This indicates that confirmation bias has an important role in our model, facilitating polarization of opinions, since agents are influenced little by peers with conflicting opinions. The effect is larger as lambda increases, i.e. as under-reaction is growing, confirmation bias becomes more effective in generating polarization. This seems to indicate that a society that changes opinion slowly is more susceptible to polarisation due to confirmation bias.    

From the point of view of average opinions, overall values stay around the middle of the opinion interval (0.5), indicating that even if consensus is on one of the two models (optimistic or pessimistic), the exact model changes from one simulation to another, due to the symmetry of the model definition. That means that the belief distribution turns out to be characterised by path-dependence. This also applies in the case of two opinion clusters, when the opinion of the larger group changes from one simulation to another.

%\begin{figure}[ht]
%\center
% \includegraphics[width=\linewidth]{figures_clipped/probabilities_evolution2.png}
%  \caption{Example of an evolution plot of the agents subjective probabilities in the baseline scenario with final polarization of the network. The simulation is realized with a fully connected network with $N=500$ agents, optimistic and pessimistic models probabilities equals to $\pi_o=0.01$ and $\pi_p=0.99$, respectively. Each agent has a degree of under-reaction  $\lambda=1.0$ and a strength of directional reasoning $\phi = 0.9$. In the figure, after about 60 iterations, a cluster of agents of the network has a final subjective probability $p = 0.99$, while the rest of the agents has the probability $p = 0.01$.  At the steady state the network reached an equilibrium with a final polarization in two different opinions. \textcolor{red}{Potrebbe essere interessante metterci anche la distribuzione delle probabilità per far vedere che non sempre la rete si divide esattamente a metà?}}
%  \label{fig:evolution_2}
%\end{figure}

% showing the emergence of polarization for a high enough $\phi$. %\textcolor{red}{To DO: add the definition of the effective number of clusters or a reference to the paper of Alina where it is defined.}

\begin{figure}[ht]
\center
\begin{subfigure}[b]{0.45\textwidth}
        \includegraphics[width=\linewidth]{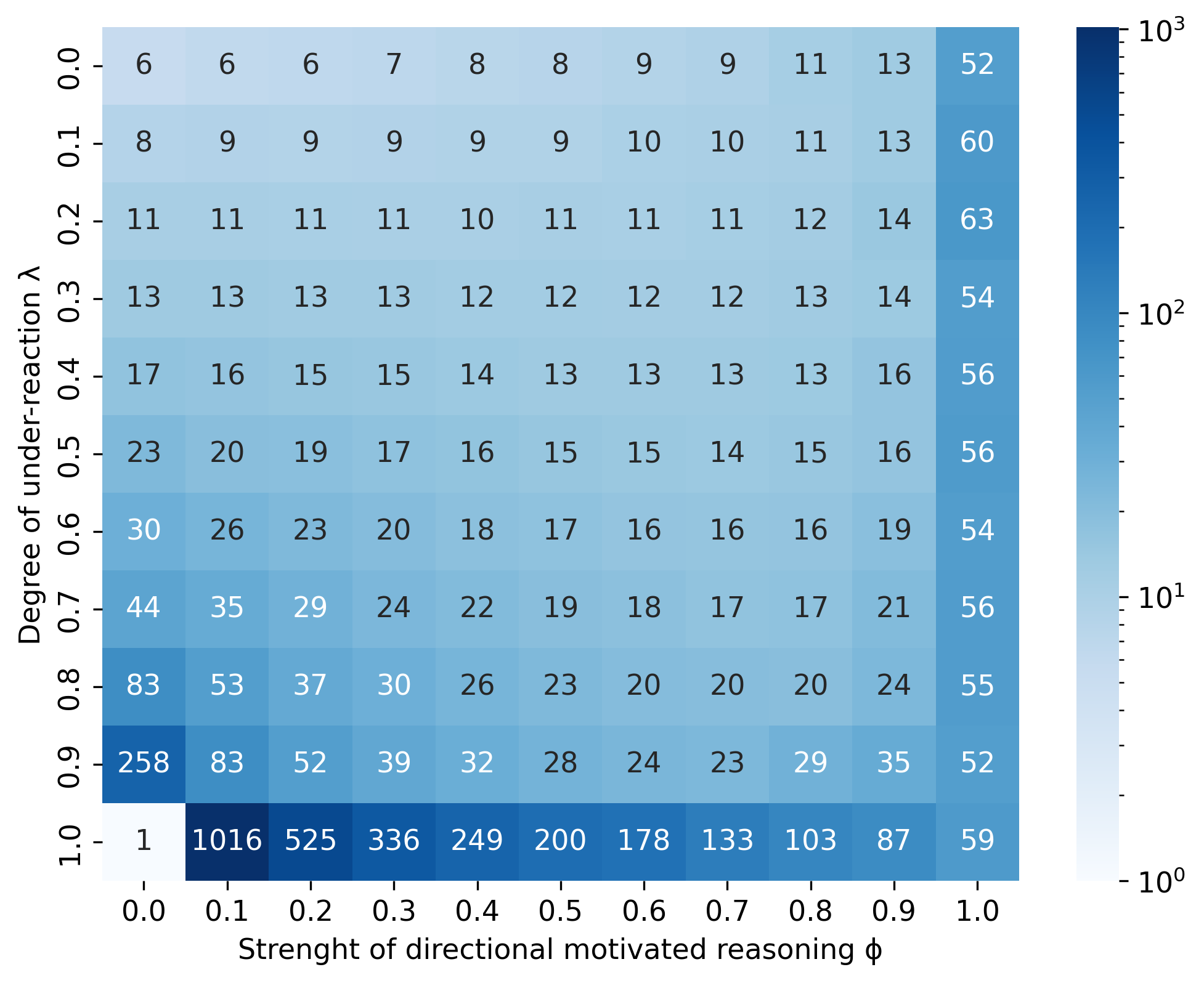}
        \caption{Whole parameter space}
    \end{subfigure}
    \hspace{0.05\textwidth}
    \begin{subfigure}[b]{0.45\textwidth}
        \includegraphics[width=\linewidth]{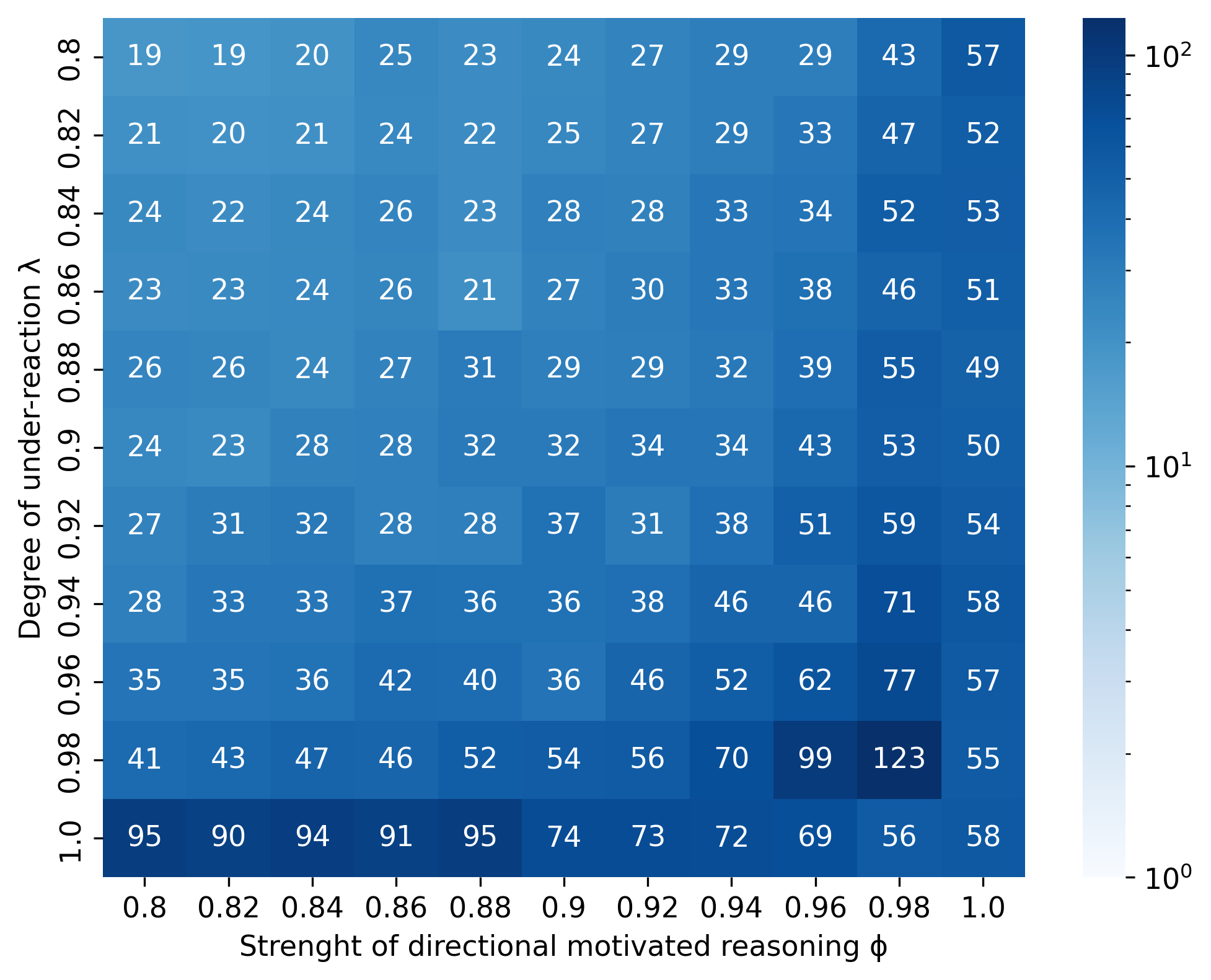}
        \caption{Zoom on the bottom right corner of the parameter space}
    \end{subfigure}
  \caption{Average number of iterations in the baseline (no-lobbyist) scenario. The average number of iterations to reach out the equilibrium in the network is represented as a function of strength of directional motivated reason $\phi$ and the degree of under-reaction $\lambda$. %Values are averaged on 150 independent runs of each setting. T
  The colorbar is in logarithmic scale.}
  \label{fig:heatmap_iterations0}
\end{figure}

Another basic property of the model is that, while the model reaches consensus generally very fast, the time required to reach the stable state is increasing in $\lambda$ (as shown in Figure \ref{fig:heatmap_iterations0}). In particular, as $\lambda$ approaches 1, nodes take a larger number of iterations to reach a stable configuration, because each of them is less sensitive to new information and under-reacts. The relation between the convergence time and $\phi$ is, however, not so linear. For low $\lambda$ values, large $\phi$ seems to slow down convergence, i.e. if the individuals do not under-react, then confirmation bias slows down convergence as society polarises. However, when $\lambda$ is large, the opposite effect is visible: since individuals under-react, polarisation is faster as confirmation bias increases: they are not attracted by peers from the opposite opinion groups and so the simulations ends faster.

\subsection*{One-lobbyist scenario} \label{Sec:Results_one_lobbyist}

In the second setting, we introduce a first lobbyist agent that randomly sends signals to the network, with the aim of shifting the distribution of subjective probabilities towards the model it is supporting (namely, towards the pessimistic model $p$).

\begin{figure}[ht!]
\center
%\begin{subfigure}[b]{0.45\textwidth}
%        \includegraphics[width=\linewidth]{figures_clipped/heatmap_probabilities_avg_average_opinions1.png}
%        \caption{Baseline scenario}
%    \end{subfigure}
\begin{subfigure}[b]{0.45\textwidth}
        \includegraphics[width=\linewidth]{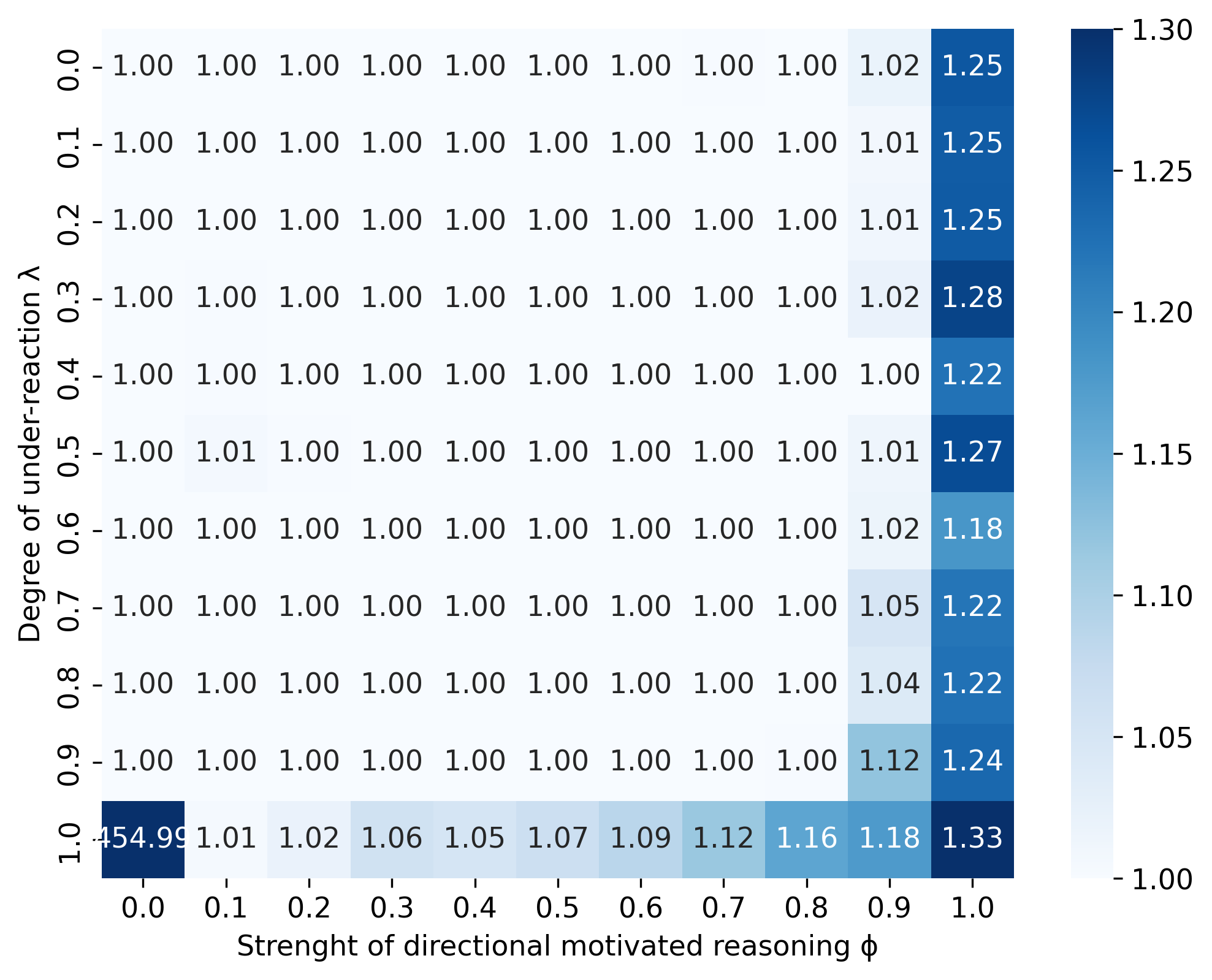}
        \caption{Average effective number of clusters.}
    \end{subfigure}
    \hspace{0.05\textwidth}
    \begin{subfigure}[b]{0.45\textwidth}
        \includegraphics[width=\linewidth]{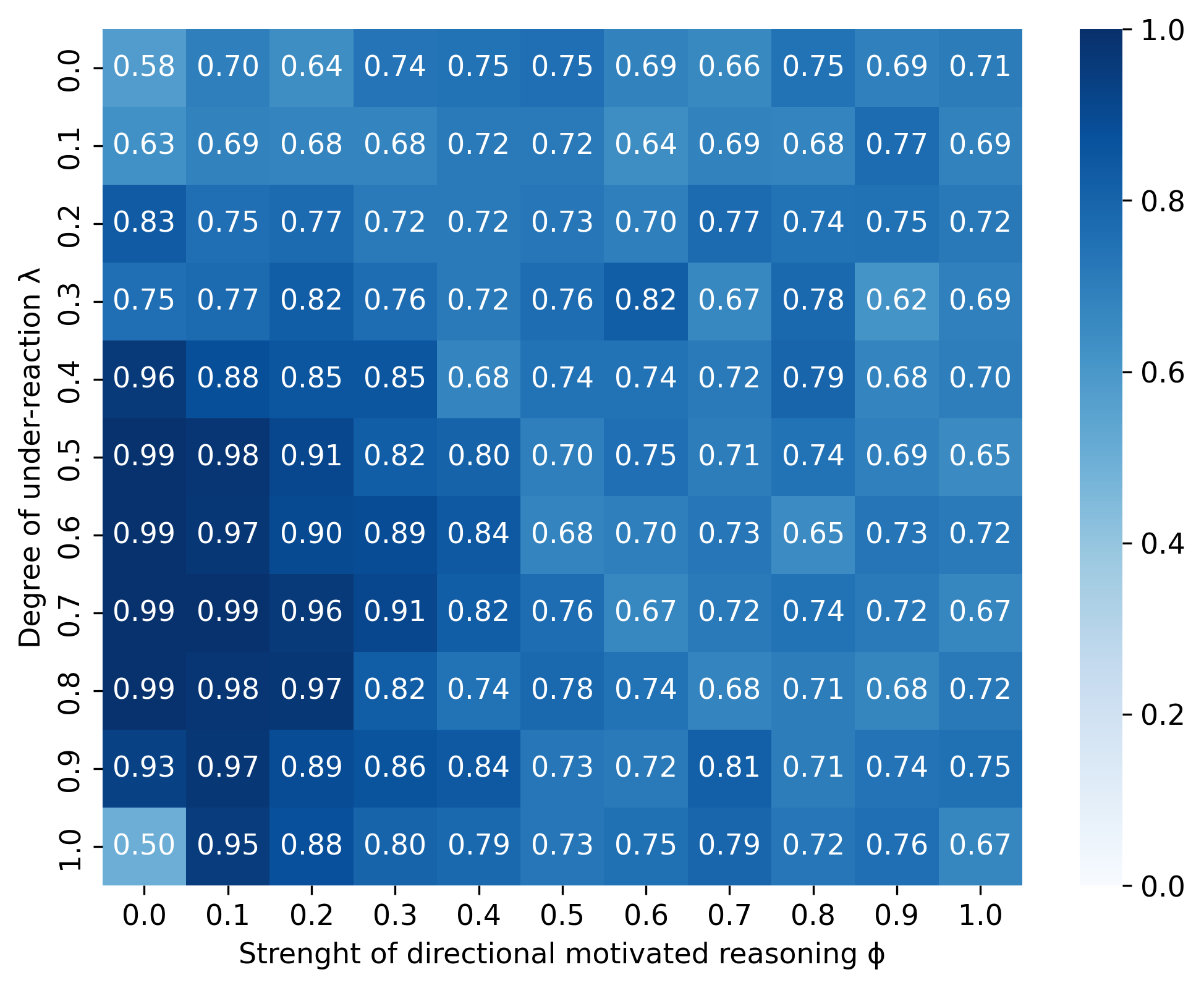}
        \caption{Average subjective probabilities.}
          \label{fig:heatmap_prob2}
    \end{subfigure}
  \caption{One-lobbyist scenario.  The effective number of clusters and average subjective probability of the final opinion distribution is represented as a function of the strength of directional motivated reason $\phi$ and the degree of under-reaction $\lambda$ in presence of a lobbyist. % Values are averaged on 150 independent runs of each setting. Simulations are performed with a fully connected network with $N=500$ agents. In the one-lobbyist scenario, t
  The lobbyists supports the pessimistic model, has a budget $B = 10,\!000$ to send its signals and can be active for a time horizon $T=100$.}
  \label{fig:heatmap_average1}
\end{figure}

In Figure \ref{fig:heatmap_average1} we show the number of clusters and  average subjective probabilities at the steady state.
While in the baseline simulation  outcomes were balanced between the two competing models (thus resulting in average subjective probabilities around 0.5 across independent runs, see Figure \ref{fig:heatmap_cluster0}), in the one-lobbyist scenario average probabilities are significantly higher over the whole parameter space, indicating that generally the lobbyist is able to steer part of the network towards supporting its own preferred model.
Furthermore, the effectiveness of the lobbyist is not globally homogeneous  over the parameter space: in fact, the average probabilities are around $0.99$ in the bottom-left corner region (indicating that, with these configurations, the lobbyist is successful in influencing the network almost always), but tend to be lower as we go farther off said region.
Moreover, we see that the lobbyist, while managing to successfully increase the average subjective probability distributions, is unable to eliminate or reduce clustering in the bottom-right corner of the map (Figure \ref{fig:heatmap_average1}).
If anything, in this region of the map an increase in polarization emerges, as the effective number of clusters increases, i.e. the size of the second cluster increases.

 The described behaviour persists independently from the size $N$ of the network, as reported in the Supplementary material where the results of the sensitivity analysis with respect to this parameter are analysed.

 These results indicate that a society with medium-high levels of under-reaction and low confirmation bias (dark blue area in Figure \ref{fig:heatmap_prob2}) are easily influenced by a lobbyist, while when confirmation bias is strong or the population is more dynamic (low under-reaction), the peer-effect can contrast the lobbyist. This can generate  consensus on the opposite opinion in a minority of simulations, when confirmation bias is low, or increase polarisation when  confirmation bias is high. There appears to be a type of phase transition at the boundary, where the lobbyist does not have full power any more.  

%\begin{figure}[ht]
%\center
%\begin{subfigure}[b]{0.45\textwidth}
%        \includegraphics[width=\linewidth]{figures_clipped/heatmap_probabilities_avg_effective_number_clusters2.png}
%        \caption{Baseline scenario}
%    \end{subfigure}
%    \hspace{0.05\textwidth}
%    \begin{subfigure}[b]{0.45\textwidth}
%        \includegraphics[width=\linewidth]{figures_clipped/heatmap_probabilities_avg_effective_number_clusters3.png}
%        \caption{One-lobbyist scenario}
%    \end{subfigure}
% \caption{Comparison of the average effective number of clusters in the baseline (no-lobbyist) scenario and the one-lobbyist scenario with a focus of the bottom-right corner of the parameter space. In the figure, the average effective number of clusters of the final opinion distribution is represented as a function of strength of directional motivated reason $\phi$ and the degree of under-reaction $\lambda$. Values are averaged on 150 independent runs of each setting. The simulations are performed with a fully connected network of $N=500$ agents. In the one-lobbyist scenario, the lobbyists supports the pessimistic model, has a budget $B = 10,\!000$ to send its signals to the agents and can be active for a time horizon $T=100$.}
% \label{fig:heatmap_cluster1}
%\end{figure}

\begin{figure}[ht!]
  \centering
  % First row
    \begin{subfigure}[b]{0.45\textwidth}
        \includegraphics[width=\linewidth]{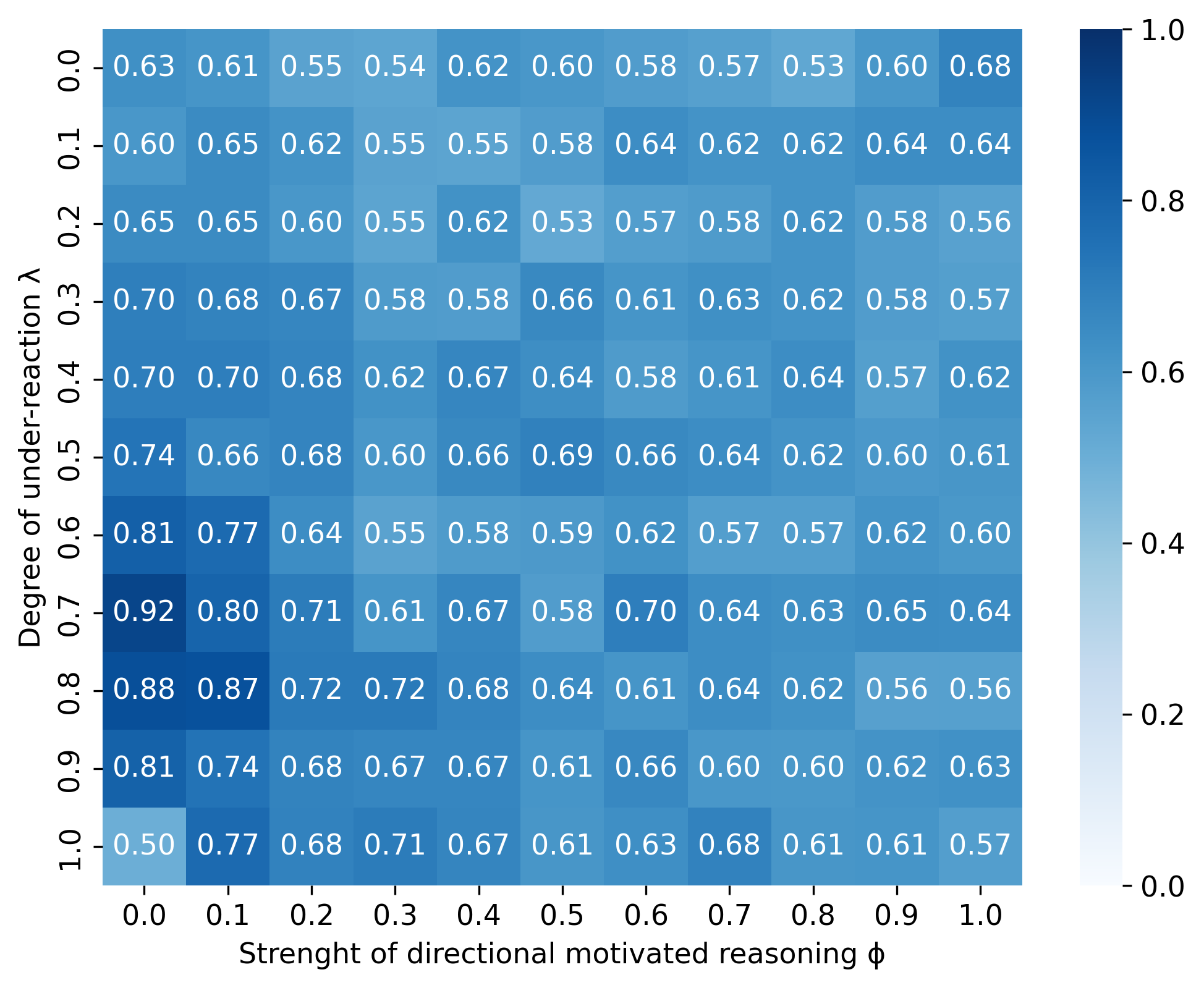}
        \caption{$B = 5,\!000$}
    \end{subfigure}
    \hspace{0.05\textwidth}
    \begin{subfigure}[b]{0.45\textwidth}
        \includegraphics[width=\linewidth]{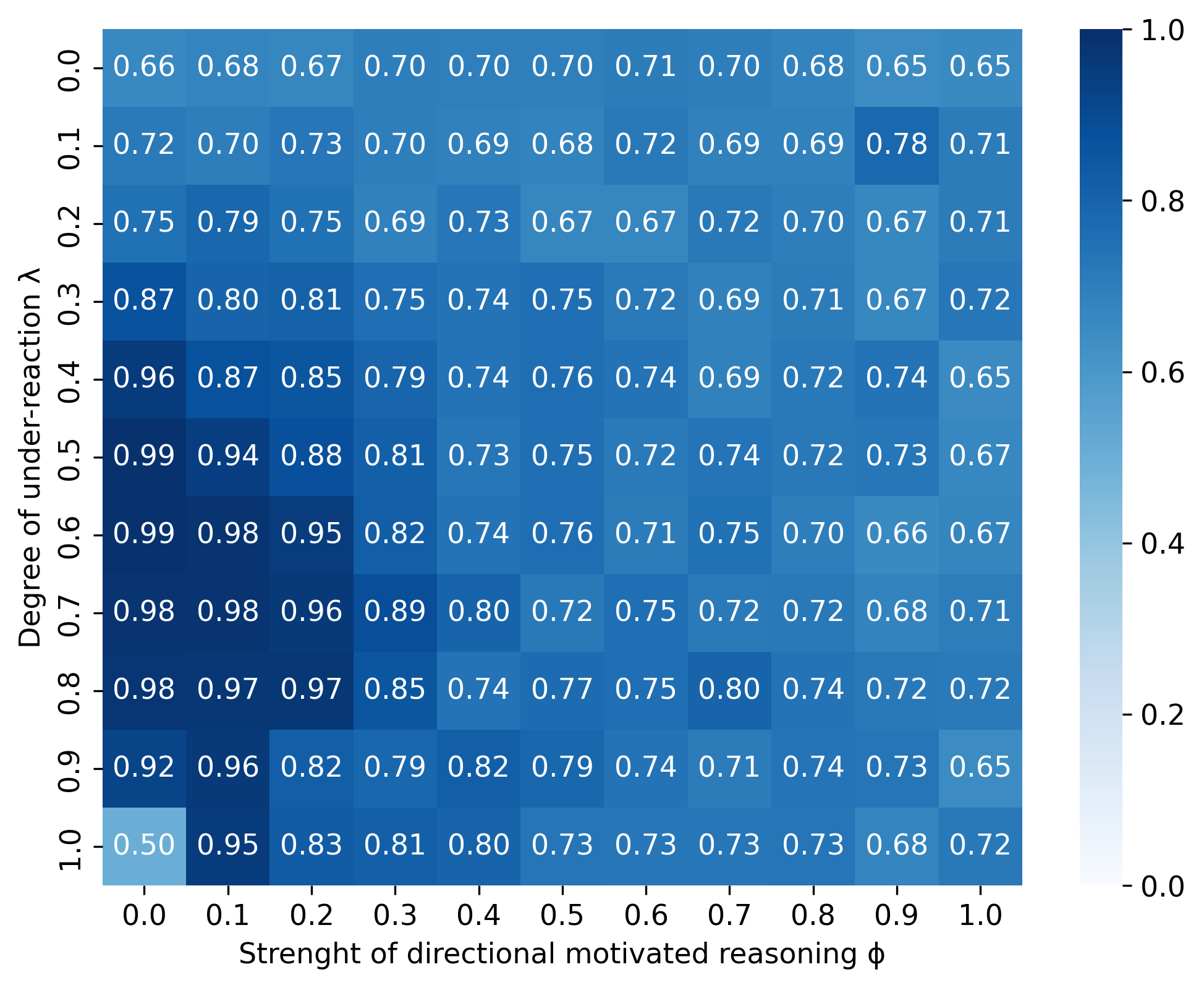}
        \caption{$B = 10,\!000$}
    \end{subfigure}

    \vspace{0em}

    % Second row
    \begin{subfigure}[b]{0.45\textwidth}
        \includegraphics[width=\linewidth]{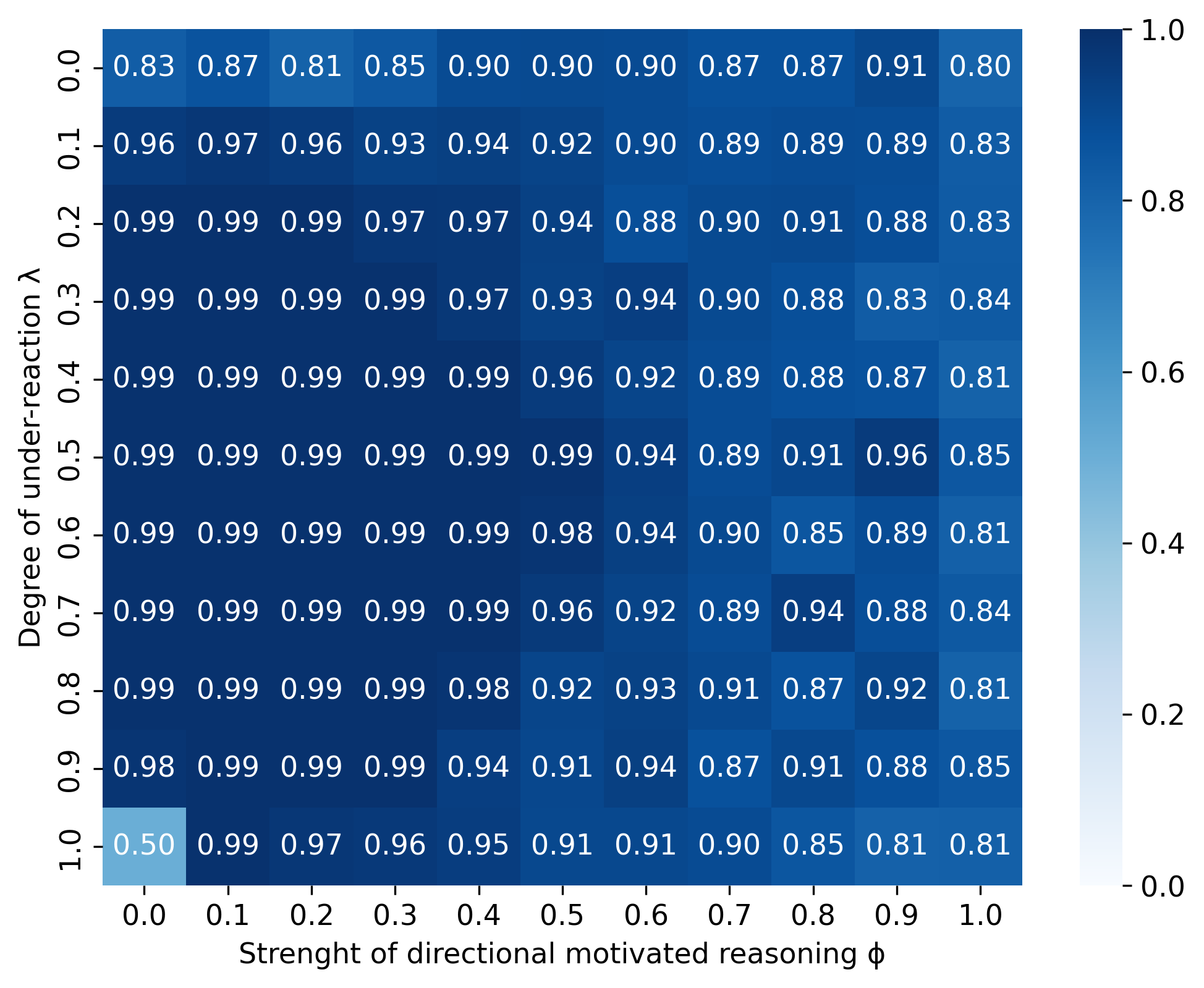}
        \caption{$B = 20,\!000$}
    \end{subfigure}
    \hspace{0.05\textwidth}
    \begin{subfigure}[b]{0.45\textwidth}
        \includegraphics[width=\linewidth]{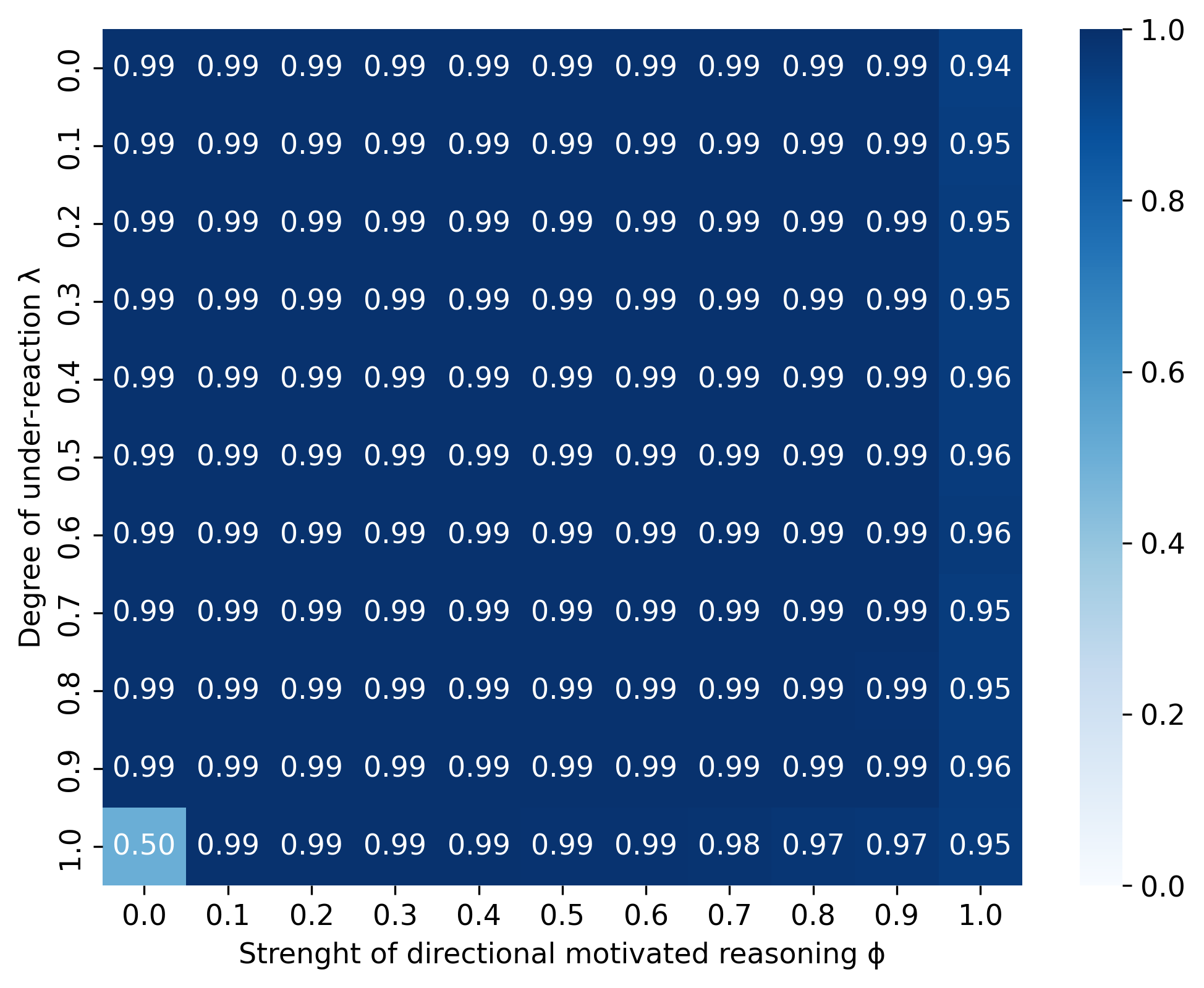}
        \caption{$B = 40,\!000$}
    \end{subfigure}
    
  \caption{Budget sensitivity analysis in one-lobbyist scenario. %Average subjective probabilities in the one-lobbyist scenario for different values of budget $B$ of the lobbyist. In the figure, t
  The average subjective probability of the final opinion distribution is represented as a function of strength of directional motivated reason $\phi$ and the degree of under-reaction $\lambda$ for different values of $B$, the budget of the lobbyist. %Values are averaged on 150 independent runs of each setting. The simulations are performed with a fully connected network of $N = 500$ agents. 
  The lobbyist supports the pessimistic model and can be active for a time horizon $T = 100$.}
  \label{fig:heatmap_budget1}
\end{figure}

To investigate further this phase transition, we performed budget sensitivity analysis, i.e. changed the budget of the lobbyist from 10,000 to 5,000, 20,000 and 40,000, allowing them to reach  10\%, 40\% and 80\% of the population, respectively. In Figure \ref{fig:heatmap_budget1}, we show the resulting average opinions. 
It is very clear that the region in which the lobbyist dominates the network (where it can impose its model almost certainly) expands linearly with the amount of resources available. With $B > 40,\!000$, the lobbyist influence is complete and global over the parameter space.
Even outside said region, however, average subjective probabilities are higher on average the higher the budget allocated to the lobbyist, indicating that the lobbyist has still a non-negligible impact on the network behaviour regardless of the specific parameter configuration.

\subsection*{Two-lobbyists scenario}

In the third scenario under consideration, we bring into the picture a second lobbyist which has the same endowments and the same kind of strategies as the first one, but that supports the competing model (namely, the ``optimist'' model).

\begin{figure}[ht!]
\center
   % \begin{subfigure}[b]{0.45\textwidth}
   %     \includegraphics[width=\linewidth]{figures_clipped/heatmap_probabilities_avg_average_opinions1.png}
   %     \caption{Baseline scenario (no lobbyists)}
   % \end{subfigure}
   \begin{subfigure}[b]{0.45\textwidth}
        \includegraphics[width=\linewidth]{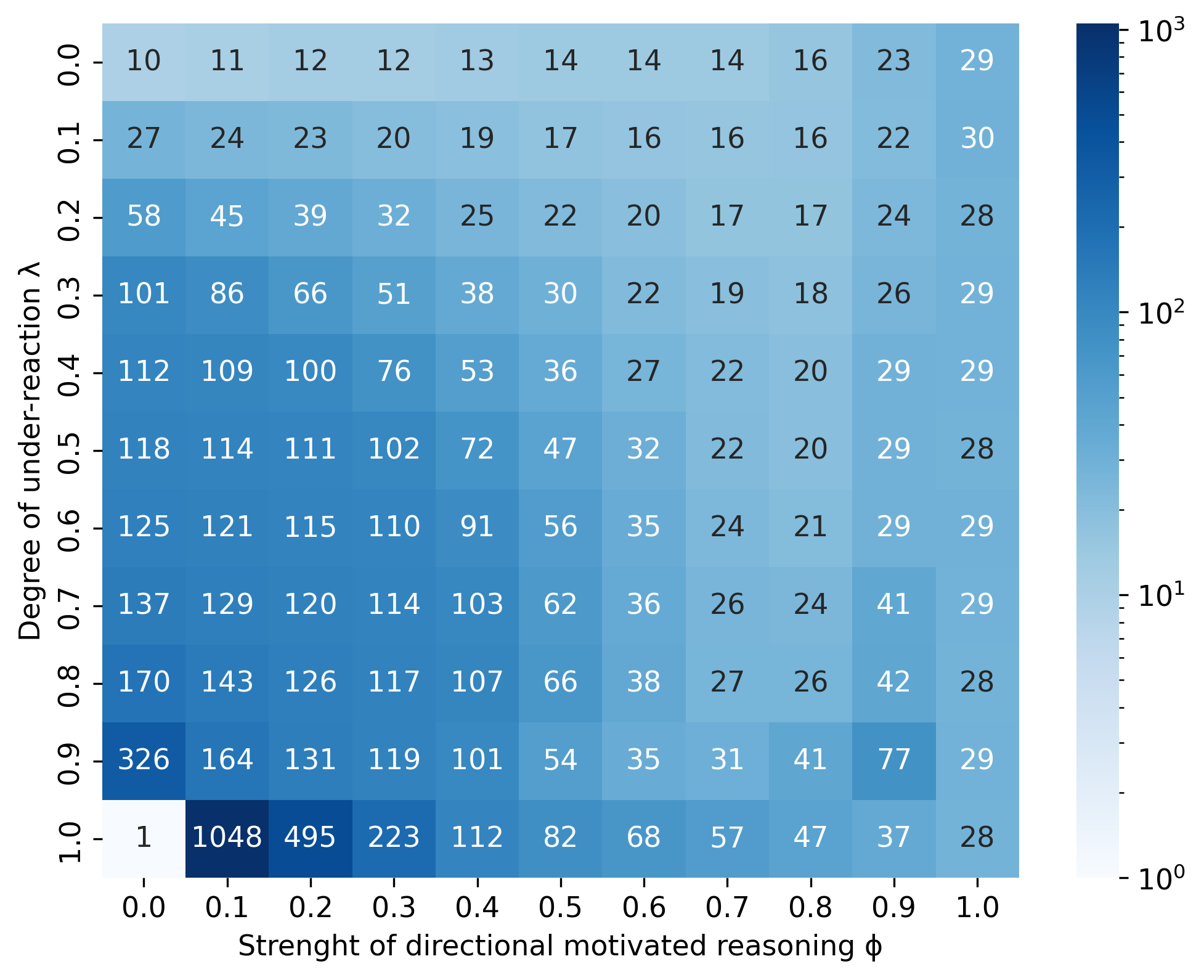}
        \caption{Average number of iterations.}
        \label{fig:heatmap_iteration2}
    \end{subfigure}
    \hspace{0.05\textwidth}
    \begin{subfigure}[b]{0.45\textwidth}
        \includegraphics[width=\linewidth]{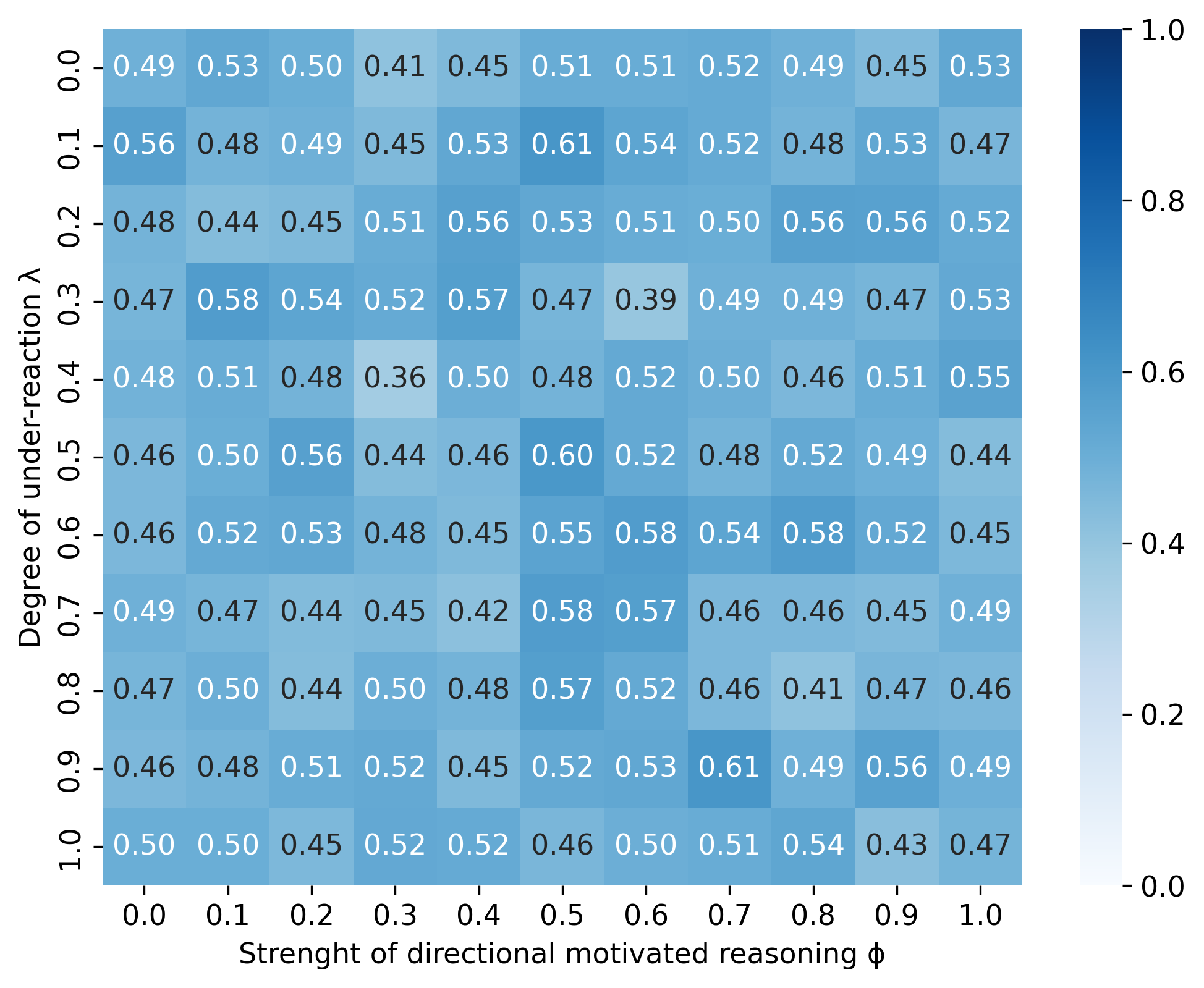}
        \caption{Average subjective probabilities.} \label{fig:heatmap_average2}
    \end{subfigure}
\begin{subfigure}[b]{0.75\textwidth}
     \includegraphics[width=\linewidth]{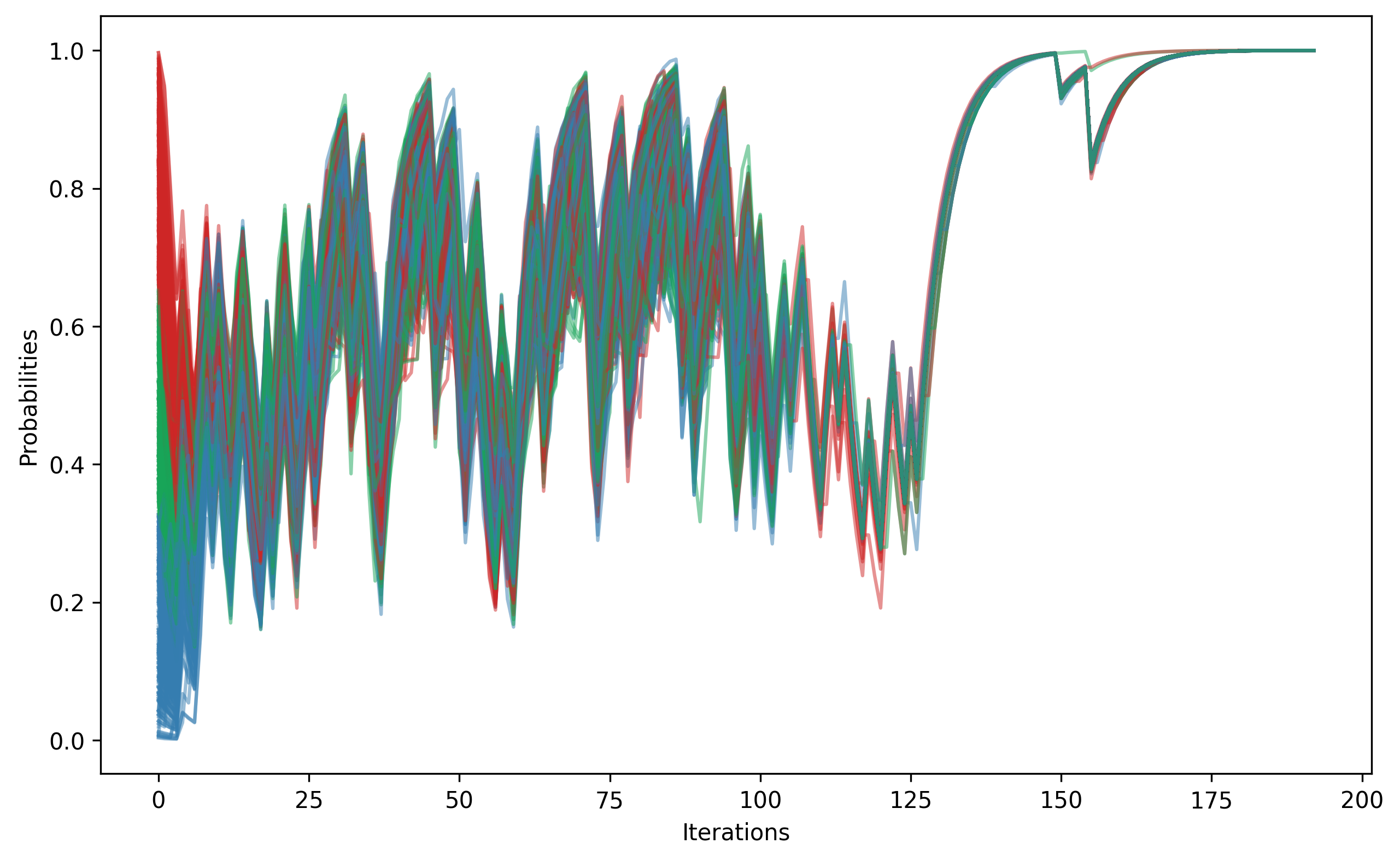}
  \caption{Example evolution plot. %The simulation is realized with a fully connected network with $N = 500$ agents, optimistic and pessimistic models probabilities equals to $\pi_o = 0.01$ and $\pi_p = 0.99$, respectively. 
  Each agent has a degree of under-reaction $\lambda = 1.0$ and a strength of directional reasoning $\phi = 0.9$. %In the simulation the lobbyists supports opposing models (one the optimistic model and the other the pessimistic one). Both of them are endowed with a budget $B = 10,\!000$ to send its signals to the agents, i.e. can reach out an average of $20\%$ of the agents in each time  steps and can be active for a time horizon $T = 100$. This means that, in this evolution plot, 
  Within the active time horizon (100 iterations), none of the lobbyists is able to definitely attract all agents opinions near it supporting models, thus the agents' opinions strongly oscillate. After that, the simulation evolves without any lobbyist action and the network quickly reaches a final consensus, as in the baseline scenario.} \label{fig:evolution_3}
  \end{subfigure}
  \caption{The two-lobbyists scenario. The average number of iterations and average subjective probabilities of the final opinion distribution are represented as a function of strength of directional motivated reason $\phi$ and the degree of under-reaction $\lambda$. %Values are averaged on 150 independent runs of each setting. The simulations are performed with a fully connected network of $N = 500$ agents. In the two-lobbyist scenario, t
  The lobbyists supports opposing models (one the optimistic model and the other the pessimistic one). Both of them are endowed with a budget $B = 10,\!000$ to send its signals to the agents, i.e. can reach out an average of $20\%$ of the agents in each time  steps and can be active for a time horizon $T = 100$.}
  \label{fig:ev_6}  
\end{figure}

In this configuration, average subjective probabilities across simulations seem to balance out, just as in the baseline scenario, as it is shown in Figure \ref{fig:heatmap_average2}.
In fact, as the two lobbyists are identical except for the specific model supported, their competing influence seems to cancel out, and the average subjective probabilities gravitates around $\bar{p} = 0.5$, regardless of the parametrization.

The main difference emerging in this case in comparison to the baseline scenario is that competition between lobbyists over influence on the network results in an increase in the number of iterations needed to reach the stable state, as shown in Figure \ref{fig:heatmap_iteration2}.

%\begin{figure}[ht]
%\center
%    \begin{subfigure}[b]{0.45\textwidth}
%        \includegraphics[width=\linewidth]{figures_clipped/heatmap_probabilities_avg_number_iterations1.png}
%        \caption{Baseline scenario (no lobbyists)}
%    \end{subfigure}
%    \hspace{0.05\textwidth}
%    \begin{subfigure}[b]{0.45\textwidth}
%        \includegraphics[width=\linewidth]{figures_clipped/heatmap_probabilities_avg_number_iterations3.png}
%        \caption{Two-lobbyists scenario}
%    \end{subfigure}
%  \caption{Comparison of the average number of iterations in the baseline (no-lobbyist) scenario and the two-lobbyists scenario. In the figure, the average number of iterations to reach out the equilibrium in the network is represented as a function of strength of directional motivated reason $\phi$ and the degree of under-reaction $\lambda$. Colorbar is in logarithmic scale and values are averaged on 150 independent runs of each setting. The simulations are performed with a fully connected network of $N = 500$ agents. In the two-lobbyist scenario, the lobbyists supports opposing models (one the optimistic model and the other the pessimistic one). Both of them are endowed with a budget $B = 10,\!000$ to send its signals to the agents, i.e. can reach out an average of $20\%$ of the agents in each time  steps and can be active for a time horizon $T = 100$.}
%  \label{fig:heatmap_iteration2}
%\end{figure}

\begin{figure}[ht!]
  \centering
   % First row
%    \begin{subfigure}[b]{0.45\textwidth}
%        \includegraphics[width=\linewidth]{figures_clipped/T1.png}
%        \caption{$T = 300$}
%    \end{subfigure}
%    \hspace{0.05\textwidth}
%    \begin{subfigure}[b]{0.45\textwidth}
%        \includegraphics[width=\linewidth]{figures_clipped/T2.png}
%        \caption{$T = 500$}
 %   \end{subfigure}

    % opinion
    \begin{subfigure}[b]{0.45\textwidth}
        \includegraphics[width=\linewidth]{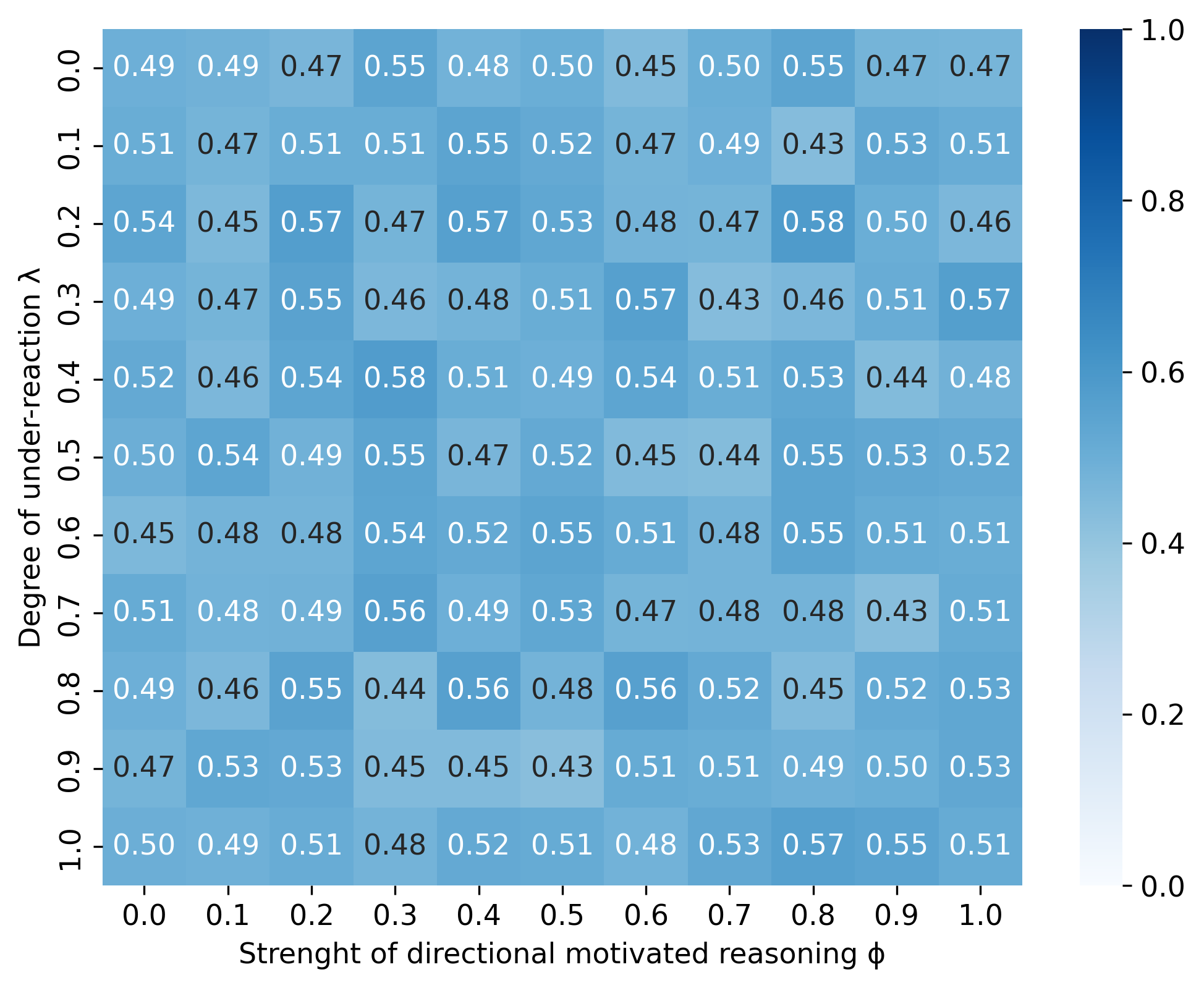}
        \caption{Average subjective probabilities for $T = 1,\!000$}
    \end{subfigure}
    \hspace{0.05\textwidth}
    \begin{subfigure}[b]{0.45\textwidth}
        \includegraphics[width=\linewidth]{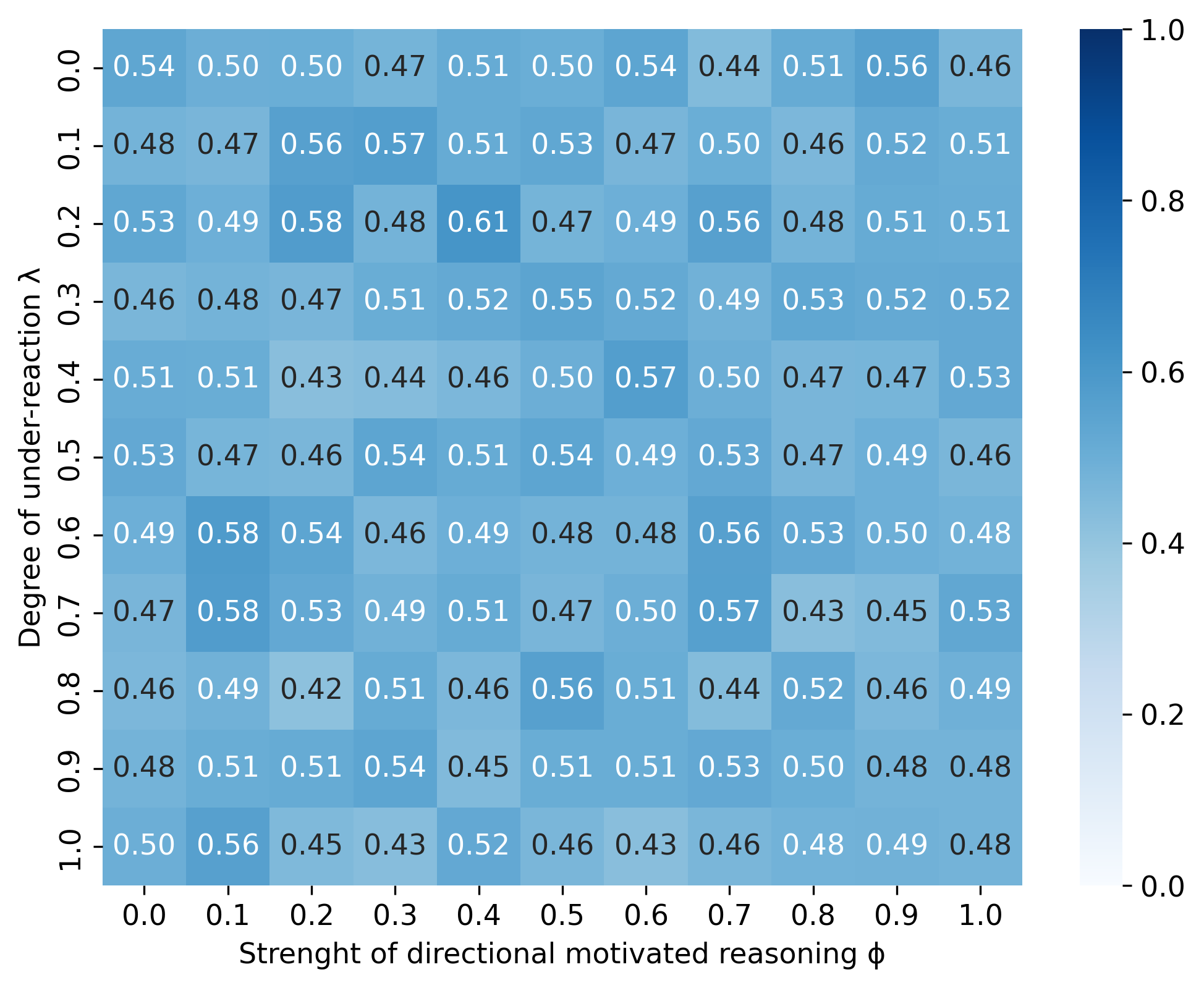}
        \caption{Average subjective probabilities for $T = 2,\!000$}
    \end{subfigure}

\vspace{0em}

 % \begin{subfigure}[b]{0.45\textwidth}
        %\includegraphics[width=\linewidth]{figures_clipped/heatmap_probabilities_avg_average_opinions_T3_ZOOM_no_numb.png}
       % \caption{Zoom of average subjective probabilities for $T = 1,\!000$}
   %\end{subfigure}
  %  \hspace{0.05\textwidth}
 %   \begin{subfigure}[b]{0.45\textwidth}
        %\includegraphics[width=\linewidth]{figures_clipped/heatmap_probabilities_avg_average_opinions_T4_ZOOM_no_numb.png}
    %    \caption{Zoom of average subjective probabilities for $T = 2,\!000$}
   % \end{subfigure}
    
 %   \vspace{0em}
    
     \begin{subfigure}[b]{0.45\textwidth}
        \includegraphics[width=\linewidth]{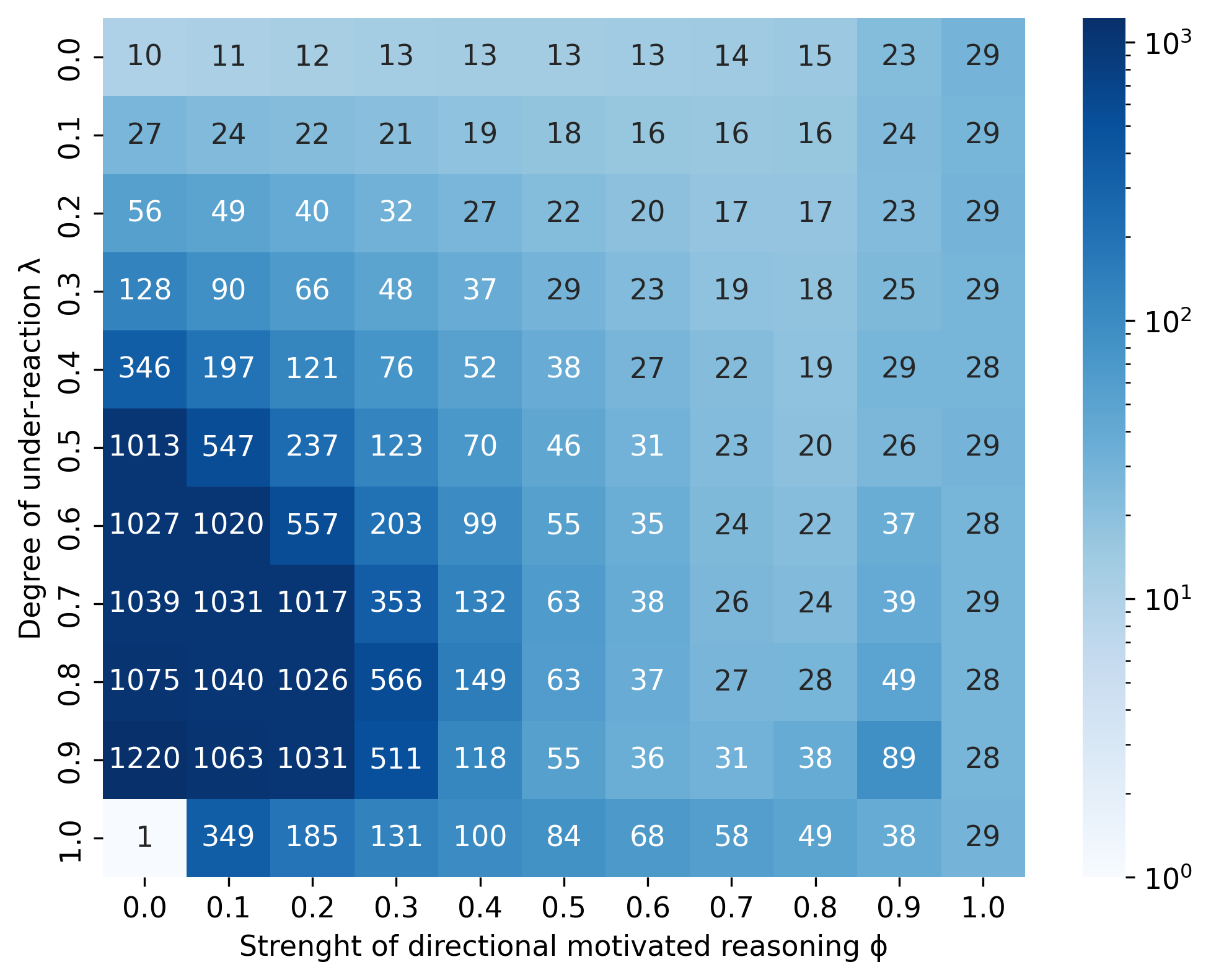}
        \caption{Average number of iterations for $T = 1,\!000$}
    \end{subfigure}
    \hspace{0.05\textwidth}
    \begin{subfigure}[b]{0.45\textwidth}
        \includegraphics[width=\linewidth]{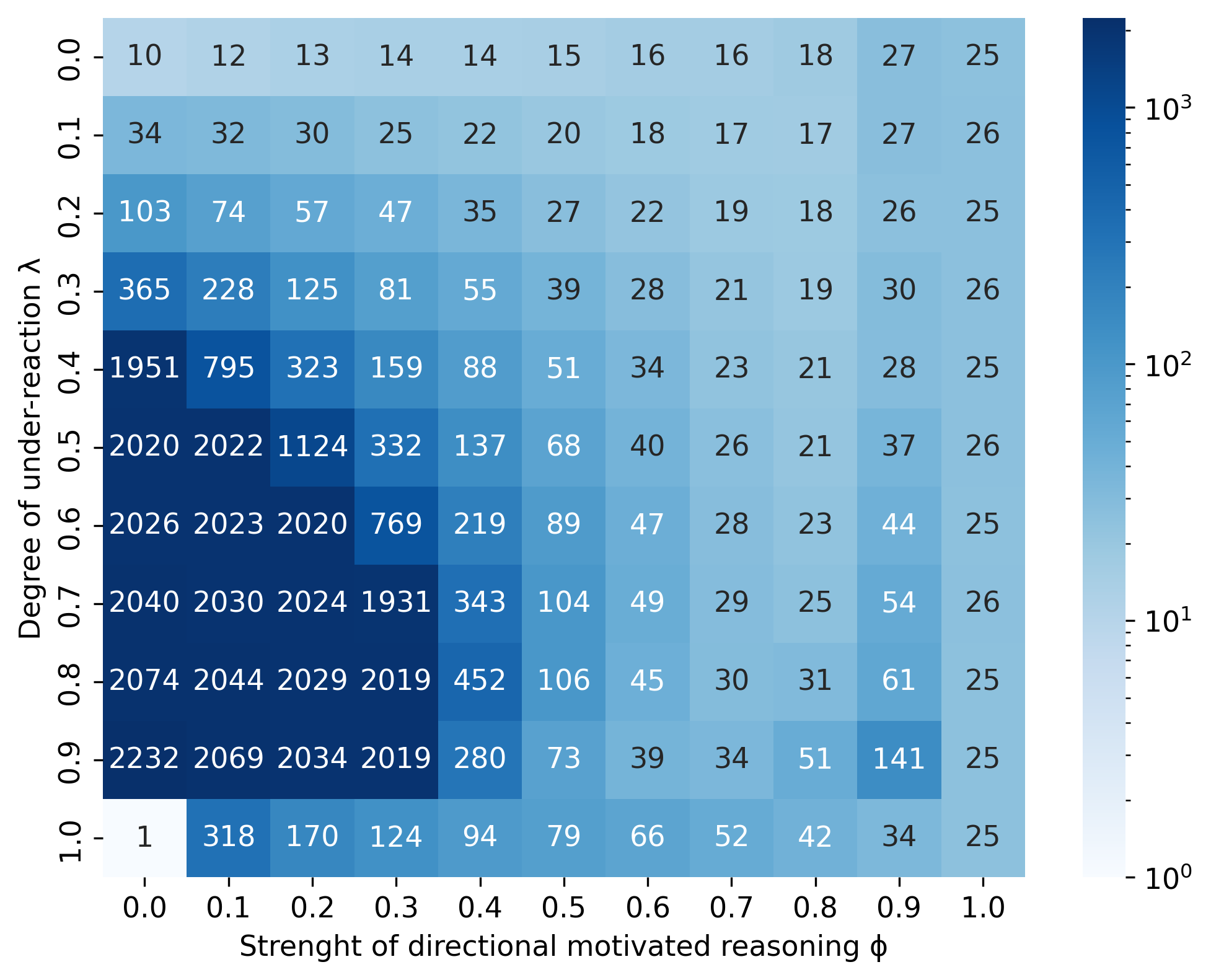}
        \caption{Average number of iterations for $T = 2,\!000$}
    \end{subfigure}

  \caption{Long lobbyist time horizons. The average subjective probability of the final opinion distribution, and the average number of iterations are represented as a function of strength of directional motivated reasoning $\phi$ and the degree of under-reaction $\lambda$ for different $T$, the maximum number of steps where lobbyists are active. %Values are averaged on 150 independent runs of each setting. The simulations are performed with a fully connected network of $N = 500$ agents; 
  The lobbyists support opposing models and have a fixed interaction rate with the agents of the networks for each step of the simulation, set at 0.2.}
  \label{fig:heatmap_T_2_opinions}
\end{figure}

In other words, the joint activity of the two lobbyists prevents the network from reaching an equilibrium as long as the lobbyists send signals; this is particularly true in the region that, in the previous scenario, was associated with higher lobbyist efficacy, strengthening the impression that, for that subset of configurations, it is easier for the external agent to project its influence on the network.
Figure \ref{fig:evolution_3} represents the evolution  of a typical run of the model in one of those configurations. Differently from what happens in the baseline scenario, where the network converges to a stable state relatively quickly, the system undergoes persistent fluctuations for as long as the two lobbyists are active on the network. Shortly after they stop emitting signals (beyond 100 iterations), however, the network quickly converges to a consensus, in a similar dynamics to what we have seen in the baseline scenario. It is worth noticing that while oscillations with stubborn influences also appear in linear models, here they emerge from motivated, signal-dependent responsiveness combined with time-staged strategic sending, yielding the regime patterns documented in Figs. \ref{fig:ev_6}-\ref{fig:heatmap_T_2_opinions}.

%\begin{figure}[ht]
%\center
% \includegraphics[width=\linewidth]{figures_clipped/probabilities_evolution3.png}
%  \caption{Example of an evolution plot of the agents subjective probabilities in the two-lobbyists scenario. The simulation is realized with a fully connected network with $N = 500$ agents, optimistic and pessimistic models probabilities equals to $\pi_o = 0.01$ and $\pi_p = 0.99$, respectively. Each agent has a degree of under-reaction $\lambda = 1.0$ and a strength of directional reasoning $\phi = 0.9$. In the simulation the lobbyists supports opposing models (one the optimistic model and the other the pessimistic one). Both of them are endowed with a budget $B = 10,\!000$ to send its signals to the agents, i.e. can reach out an average of $20\%$ of the agents in each time  steps and can be active for a time horizon $T = 100$. This means that, in this evolution plot, within the time horizon (100 iterations), none of the lobbyists is able to definitely attract all agents opinions near it supporting models, thus the agents' opinions strongly oscillate. After that, the simulation evolves without any lobbyist action and the network quickly reaches a final consensus, as in the baseline scenario.}
%  \label{fig:evolution_3}
%\end{figure}

In this last scenario, we try to go a step further and we perform a sensitivity analysis on the lobbyists' strategic horizon $T$ with the aim of investigating how this parameter affects the model's properties.
Specifically, we run several different batches of simulations by increasing the parameter $T$ to %300, 500, 
1,000 and 2,000. At the same time, we adjust the lobbyists' budget $B$ accordingly, in order to maintain the same rate of signals per time-period of 20\%, thus keeping the results comparable across specifications.
As shown in Figure \ref{fig:heatmap_T_2_opinions}, with a longer time horizon generally average opinion $\bar{p}$ remains about 0.5 in most configurations, suggesting that a longer horizon by the lobbyists does not radically change the outcome of the simulations. The average number of iterations, instead, tends to increase linearly with $T$ in the bottom-left corner of the parameter space, the region where the lobbyists are most effective , while it does not seem to change much in the other regions, that converge very fast due to peer effects.

\paragraph*{Comparing Lobbyists' performance across different timing strategies}

As a final exercise within the two-lobbyists scenario, we set up an experiment with the aim to investigate whether timing in the lobbyist strategies has a relevant impact on the outcome of the simulations. While insofar lobbyists have been assumed to spread out their signals uniformly over time, now we define two additional possible strategies, which we call "frontloading", and "backloading".
In lobbying, frontloading means concentrating influence efforts at the earliest stages of the process.
In the model, this is operationalized by the adoption, by the lobbyist agent, of a strategy matrix in which variable indicators for signals are concentrated in the first 20 (out of 100) iteration periods, in which the lobbyist exhausts its budget.
Conversely, backloading implies holding back resources until later stages, in order to concentrate the effort closer to the steady state. This means, in our model, that the lobbyist adopts a strategy matrix in which variable indicators for signals are concentrated in the last 20 iteration periods.

For our experiment, we set up three different scenarios: in Scenario 1, both the pessimist and the optimist lobbyists adopt a frontloading strategy; in Scenario 2, the pessimist lobbyist adopts a frontloading strategy, while the optimist sticks to a random uniform strategy; in Scenario 3, the pessimist lobbyist adopts a backloading strategy, while the optimist keeps using a random uniform strategy. We run 100 Monte Carlo simulations for each parameter configuration and for each Scenario to assess the performance of the lobbyists under these different conditions.

Results are presented in Figure~\ref{fig:heatmap_cluster_three} (see also Figures~S3 and S4 in Supplementary material). Expectedly, in the completely symmetric Scenario 1 (Figure \ref{fig:both_front}), average opinion outcomes are balanced across the parameter space, with neither lobbyist prevailing in the steady state across Monte Carlo simulations.
The picture changes starkly in Scenario 2 (Figure \ref{fig:pes_front}). In this case, as earlier in the model, we see two distinct regions emerge in the parameter space. In the region characterized by lower under-reaction (or high confirmation bias), where beliefs by nodes are more flexible and the peer effect is stronger, the frontloading (pessimist) lobbyist prevails over his opponent in almost all of the simulations. Conversely, the frontloading strategy seems to be defeated by the random uniform optimist lobbyist when under-reaction is higher and motivated reasoning component is lower.
In contrast, in Scenario 3 (Fig \ref{fig:pes_back}), the situation is reversed: the pessimist lobbyist (now backloading) dominates in the high-$\lambda$, low-$\phi$ configurations, but is dominated by the optimist elsewhere.

This behaviour can be readily interpreted by considering the core mechanisms driving the evolution of the network over time in our model.
Frontloading is particularly effective when agents are highly responsive to new information and early signals can trigger reinforcing dynamics, such as cascades or clustering effects. By influencing opinions early, lobbyists can increase the likelihood that the system evolves toward their preferred equilibrium and can potentially reduce the need for later interventions.

Backloading, by contrast, targets the final opinion distribution directly, which is often the variable of interest for the lobbyist. Backloading is advantageous when agents under-react to signals or when early fluctuations are likely to reverse over time.
By concentrating resources late, lobbyists maximize the impact of each signal on the final state, minimizing wasted influence and counteracting opposing signals. Backloading can be particularly effective in competitive environments where multiple lobbyists attempt to steer opinions, as late interventions can override prior influence and shift the final outcome.

Interestingly, the regions of parameter space where time-dependent strategies outperform the uniform strategy roughly match the regions in the single-lobbyist scenario where lobbyist influence is stronger (backloading prevails) and where peer effects are stronger (frontloading prevails).

\begin{figure}[ht!]
\centering
\begin{subfigure}[b]{0.45\textwidth}
    \includegraphics[width=\linewidth]{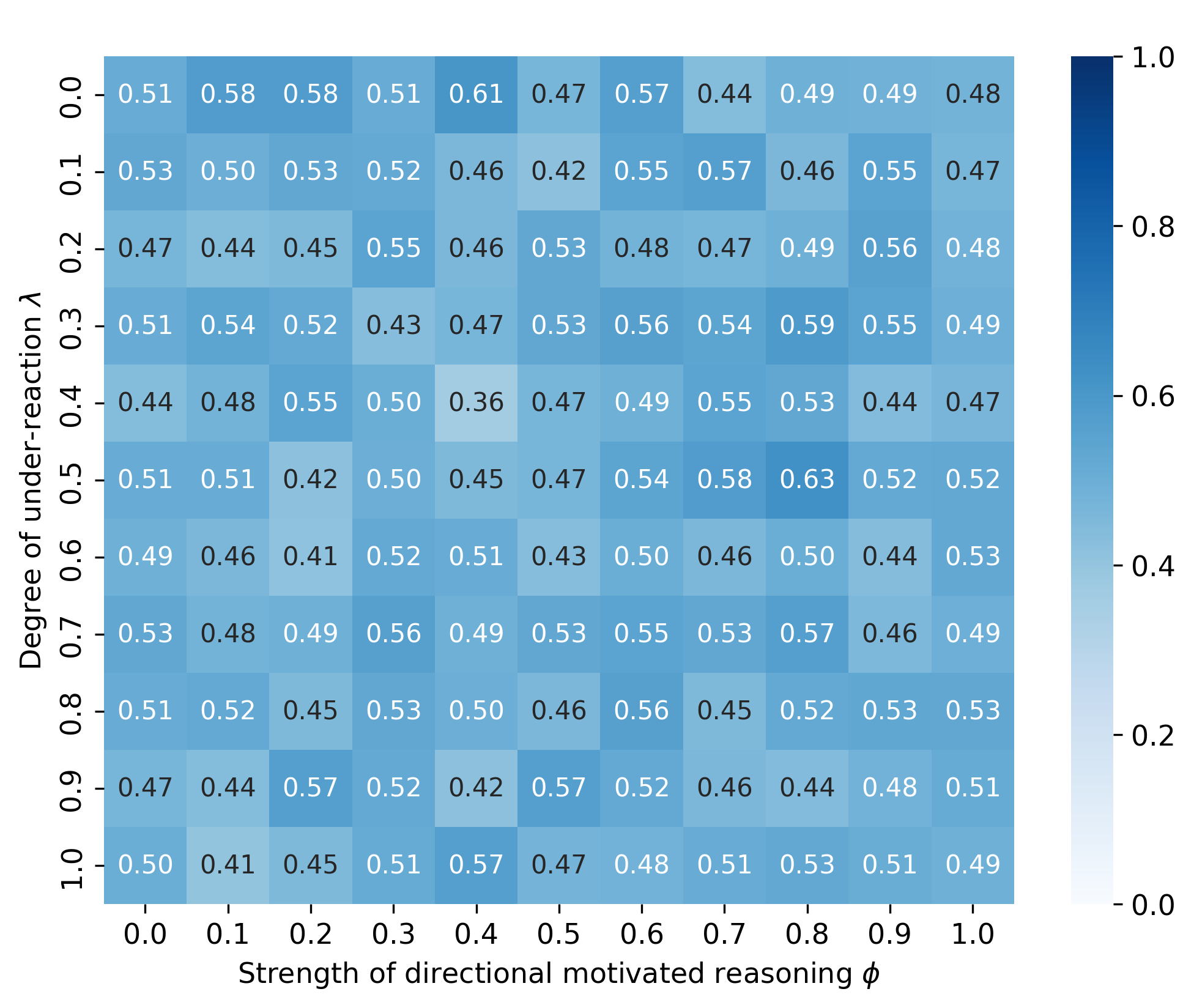}
    \caption{Scenario 1}
    \label{fig:both_front}
\end{subfigure}
\hspace{0.03\textwidth}
\begin{subfigure}[b]{0.45\textwidth}
    \includegraphics[width=\linewidth]{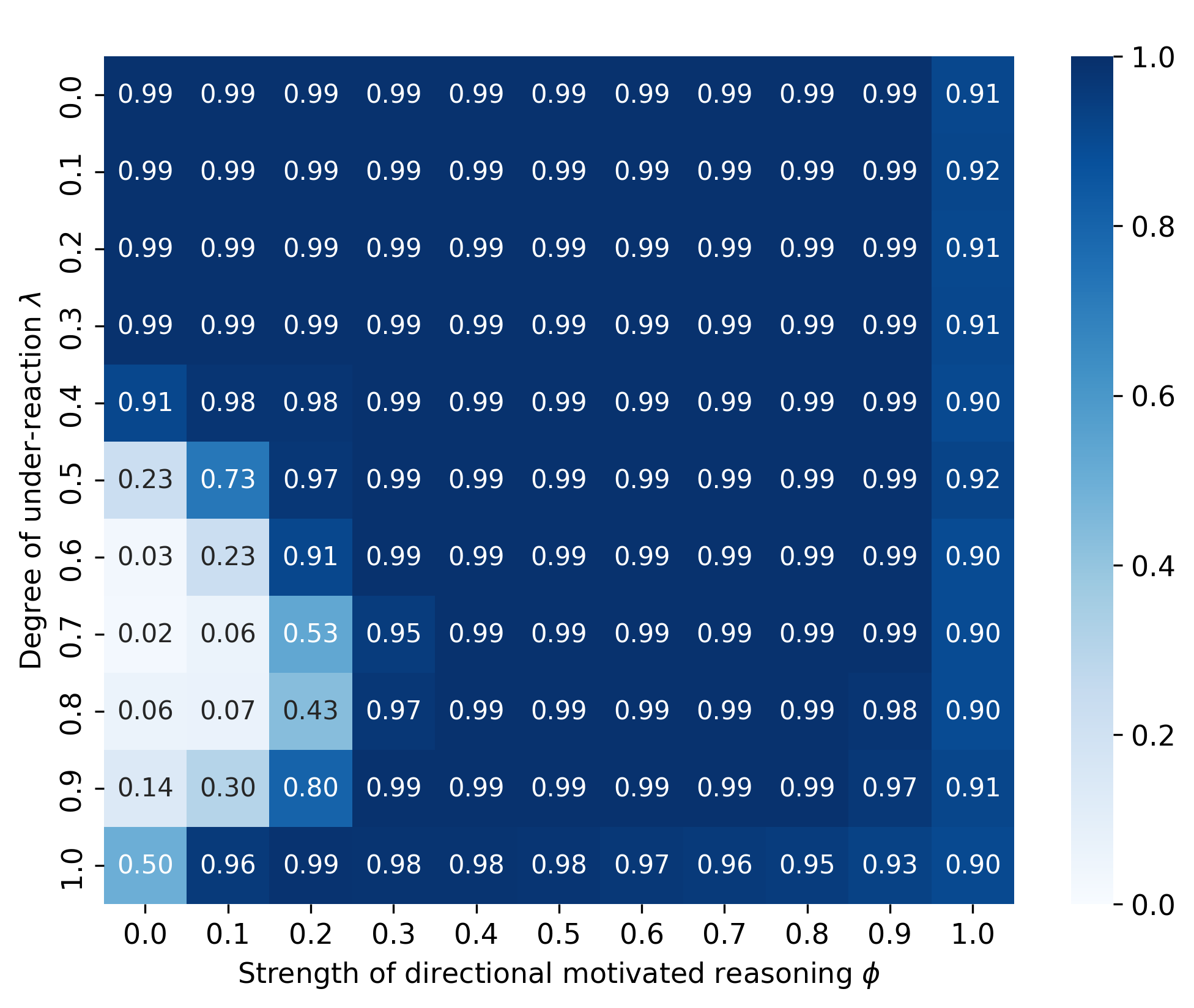}
    \caption{Scenario 2}
    \label{fig:pes_front}
\end{subfigure}
\hspace{0.03\textwidth}
\begin{subfigure}[b]{0.45\textwidth}
    \includegraphics[width=\linewidth]{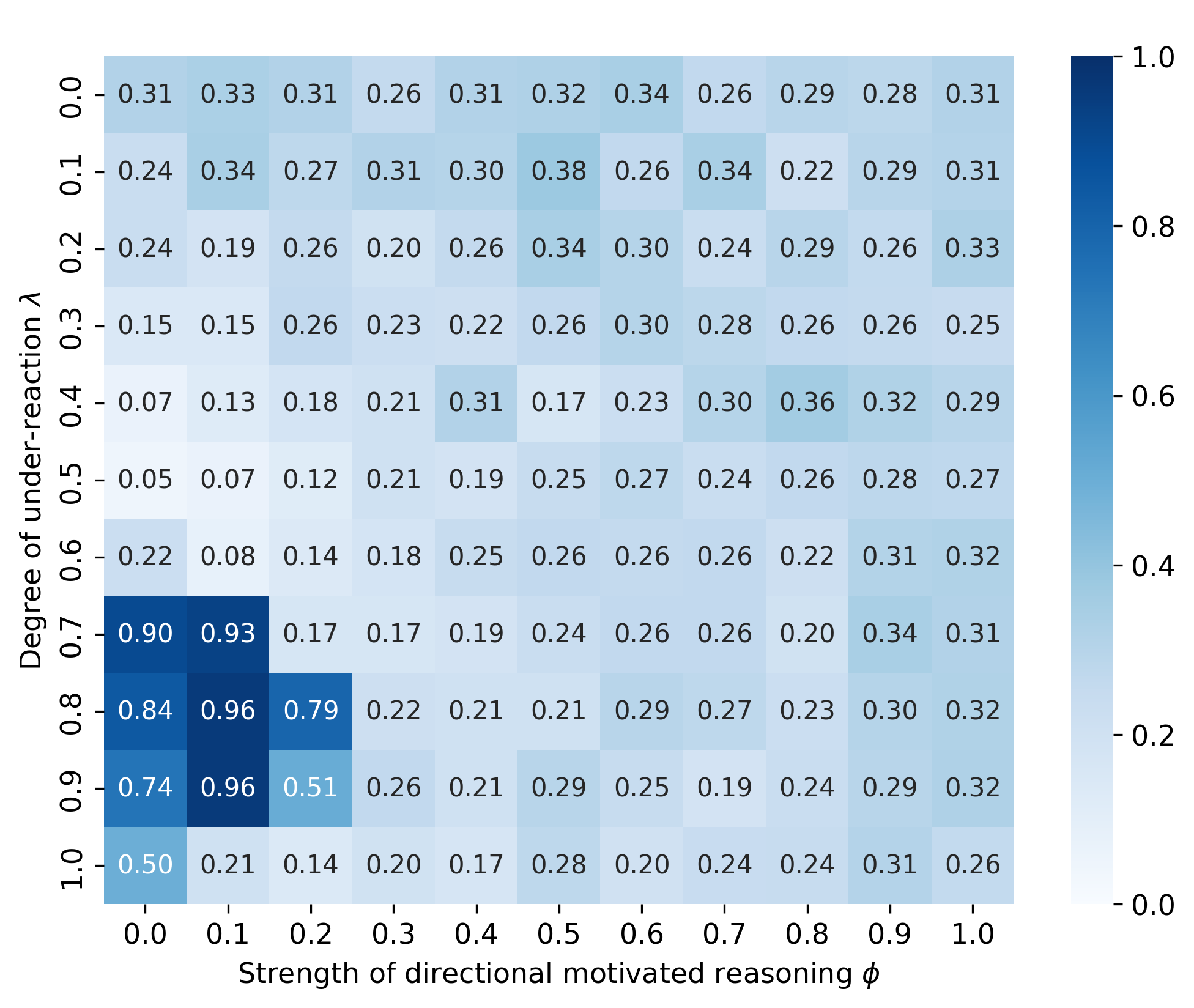}
    \caption{Scenario 3}
    \label{fig:pes_back}
\end{subfigure}
\caption{Average steady-state subjective probabilities under different lobbying timing strategies. Panel (a) shows outcomes when both lobbyists adopt a frontloading strategy. Panel (b) depicts the case where the pessimist lobbyist frontloads while the optimist follows a uniform strategy. Panel (c) illustrates the reverse scenario with the pessimist backloading.
Values are averaged over 100 independent runs for each setting. Simulations are performed on a fully connected network of $N=500$ agents and endowing each lobbyist with a budget $B=10,000$.}
\label{fig:heatmap_cluster_three}
\end{figure}

\section*{Discussion}

%The Discussion should be succinct and must not contain subheadings.
In this work, we introduce a model of opinion formation in which agents update beliefs via a Bayesian-like rule in a social-learning setting, incorporating cognitive biases such as under-reaction and directionally motivated reasoning (confirmation bias). The model includes lobbyists that are characterised by strategies (a set of agents to target at each iteration) and are budget constrained. In line with our scope, we study lobbying as an opinion-influence mechanism; the policy stage is exogenous and can be connected via a simple mapping without altering the core dynamics. After having provided some game theoretic examples, we studied the model numerically and observed the emergence of two polarised clusters when confirmation bias is large,  even in the absence of lobbying. One lobbyist with an uniform strategy can attract the entire population to the opinion it promotes if the budget is large enough or if the individuals have low confirmation bias and medium under-reaction levels. Otherwise, a polarised minority continues to coexist with the cluster that agrees with the lobbyist. In the presence of two opposing lobbyists, the final steady states obtained are similar to those without lobbying, just that convergence happens only after the lobbyists stop influencing the network. Our oscillations are thus mechanistically distinct from those in linear stubborn-agent frameworks \cite{acemouglu2013opinion}: they require (i) nonlinearity in responsiveness (equation \eqref{eq:lambda}) and (ii) endogenous, budget-constrained sending that injects temporally structured shocks (equations \eqref{eq:wl}-\eqref{eq:w1}). As long as they are active, the population continues to oscillate between different opinions without converging. This suggests that continuous lobbying can preclude consensus or convergence.     

Our model shares some similarities with classical models of opinion dynamics, however exact mechanisms are different. The internal dynamics where an internal continuous weight is updated at discrete time steps is similar to bounded confidence models such as the Deffuant-Weisbuch~\cite{deffuant2000mixing} or Hegselman-Krause~\cite{hegselmann2002opinion} models. However, here, agents do not share the internal state with peers, but an external discrete opinion, more like in the CODA model~\cite{martins2008continuous}. Directional motivated reasoning, where agents react more strongly to opinions that are similar to their own, has a rationale similar to bounded confidence in the above mentioned models, however in our model the effect is continuous rather than based on a discrete boundary. Indeed, a large $\phi$ leads to lack of consensus, similar to bounded confidence models, but we only obtain two clusters. Moreover, when consensus is reached, it usually stays on one of the two opinions, with all internal weights close to the range limit, not in the middle like bounded confidence models. This is more realistic, and models individuals who actually take a decision in the end, rather than remaining undecided and holding a moderate opinion $\sim 0.5$. In this regard, our model is more similar to the voter model~\cite{clifford1973model} or to other discrete models. 

Lobbyists are an important element of novelty in our model, since they are represented as more than an external field \cite{das2014modeling,li2020effect, sirbu2017opinion,sirbu2013opinion, pansanella2023mass} or zealot/stubborn agent\cite{mobilia2013commitment,acemouglu2013opinion}. Specifically, their effect on other agents can vary in time (at every iteration a subset of agents is reached) and is limited by a budget. This allows to model complex strategies and scenarios without changing the baseline model definition. We have shown that with this definition, one lobbyist attracts the complete population only in a certain area of the parameter space, \emph{a lobbyist influence area}, where the population has low confirmation bias and medium levels of under-reaction. Polarisation appears as the population is more flexible (low under-reaction) or has more confirmation bias, where the \emph{peer effect} can counteract the lobbyist influence. This polarising effect has also been seen in other models where extreme external information can cause minority extremist groups~\cite{sirbu2017opinion,sirbu2013opinion}.  A phase transition seems to appear at the boundary between these two regimes, i.e. lobbyist-influence area versus peer-effect area.  When two symmetric opposing lobbyists are present, a rich behaviour is observed. During the active time horizon of the lobbyists, inside the lobbyist influence area, we observe fluctuations of opinions between the two extreme opinions, with convergence as soon as the lobbyists become silent.   
Finally, our analysis of alternative timing strategies shows that the distribution of signals is not neutral in general: frontloading can be highly effective when agents are responsive and cascades amplify early interventions, while backloading dominates when agents under-react and late interventions can directly shape the final steady state. This demonstrates that the temporal structure of lobbying is itself an additional strategic dimension of influence, and that its effectiveness depends to some degree on the cognitive bias of the population.

Future work will explore more complex lobbying strategies extracted from real data\cite{errichiello2025navigating}. We will examine how these strategies perform across the two parameter regimes and run additional simulations on more realistic social networks (e.g., scale-free graphs) that better approximate real cases. We also plan to integrate the model with statistical model-checking environments, such as MultiVeStA\cite{vandin2022automated,sebastio2013multivesta}, to gain further insight into its properties and dynamics. This paper is a first step toward the broader goal of fully characterizing the complex strategic-interaction environment generated by rational lobbyists seeking to attract consensus on their positions.

\section*{Code availability}
The presented mathematical model with behavioural bias was implemented in Python using the NDlib library \cite{rossetti2018ndlib} and is available in the GitHub repository:  \href{https://github.com/ALMONDO-Project/ALMONDO-Model}{https://github.com/ALMONDO-Project/ALMONDO-Model}. Simulations can be reproduced using the code in the same repository.  

%\textcolor{red}{In the same repository the code of the mathematical model is also available.}

%\textcolor{red}{Oppure diciamo: The model used to conduct our experiments is available, under the name AlmondoModel in the NDlib  Python library???: \href{http://ndlib.rtfd.io}{http://ndlib.rtfd.io}}.

\section*{Data availability}
We do not analyse or generate any datasets, because our work proceeds within a theoretical and mathematical approach. The data supporting the findings of this study are available within the paper and its supplementary information file.

\bibliography{lobby}

\section*{Acknowledgements}
The authors are grateful to two anonymous reviewers for their insightful feedback and helpful recommendations.
\hfill

\section*{Funding Declaration}
This study received funding from the European Union - Next-GenerationEU - National Recovery and Resilience Plan (NRRP) – MISSION 4 COMPONENT 2, INVESTMENT N. 1.1, CALL PRIN 2022 PNRR D.D. 1409 14-09-2022 – ALMONDO Project, CUP N. J53D23015400001.

\section*{Author contributions statement}
%\textcolor{red}{Must include all authors, identified by initials, for example:
%A.A. conceived the experiment(s),  A.A. and B.A. conducted the experiment(s), C.A. and D.A. analysed the results.  All authors reviewed the manuscript. }

D.G. conceived the mathematical model and V.P. implemented it in NDlib Python library. All authors conceived the experiments to explore its properties. V.D.R and L.C. performed
the simulations. All authors analysed and interpreted the results. D.G., L.P., L.C. and A.S. wrote the first draft of the manuscript. All authors reviewed and approved the manuscript.

\section*{Additional information}

\paragraph{Supplementary Information} The online version will contain supplementary material % available at XXX .

%To include, in this order: \textbf{Accession codes} (where applicable); \textbf{Competing interests} (mandatory statement). 

\paragraph{Competing interests} The authors declare no competing interests.

%The corresponding author is responsible for submitting a \href{http://www.nature.com/srep/policies/index.html#competing}{competing interests statement} on behalf of all authors of the paper. This statement must be included in the submitted article file.

%\appendix 

\end{document}